\documentclass[aps,superscriptaddress, twocolumn, showpacs, a4paper, UTF8]{revtex4}
\usepackage[utf8]{inputenc}
\usepackage[T1]{fontenc}
\usepackage[colorlinks=true,linkcolor=black,citecolor=blue,urlcolor=blue]{hyperref}
\usepackage{url}
\usepackage{booktabs}
\usepackage{graphicx}    
\usepackage[online,flushleft]{threeparttable}
\usepackage{amsmath,amssymb,mathrsfs,amsfonts,nicefrac}
\usepackage{mathtools}
\usepackage{ulem}
\usepackage{dcolumn}    
\usepackage{multirow}
\usepackage{xcolor}

\definecolor{webgreen}{HTML}{008B00}

\begin{document}
\title{Investigation of Langdon effect on the nonlinear evolution of SRS from the early-stage inflation to the late-stage development of secondary instabilities}

\author{Jie Qiu}
\affiliation{Institute of Applied Physics and Computational
Mathematics, Beijing, 100094, China}
\author{Liang Hao \footnote{Corresponding author: hao\_liang@iapcm.ac.cn}}
\affiliation{Institute of Applied Physics and Computational
Mathematics, Beijing, 100094, China}
\author{Lihua Cao}
\affiliation{Institute of Applied Physics and Computational
Mathematics, Beijing, 100094, China} 
\author{Shiyang Zou}
\affiliation{Institute of Applied Physics and Computational
Mathematics, Beijing, 100094, China}

\begin{abstract}
In a laser-irradiated plasma, the Langdon effect can result in a
super-Gaussian electron energy distribution function (EEDF),
imposing significant influences on the stimulated backward Raman
scattering (SRS). In this work, the influence of a super-Gaussian
EEDF on the nonlinear evolution of SRS is investigated by three wave
model simulation and Vlasov-Maxwell simulation for plasma parameters
covering a wide range of $k\lambda_{\rm De}$ from 0.19 to 0.48 at
both high and low intensity laser drives. In the early-stage
of SRS evolution, it is found that besides the kinetic effects due
to electron trapping [Phys. Plasmas 25, 100702 (2018)], the Langdon
effect can also significantly widen the parameter range for the
absolute growth of SRS, and the time for the absolute SRS to reach
saturation is greatly shorten by Langdon effect within certain
parameter region. In the late-stage of SRS, when secondary
instabilities such as decay of the electron plasma wave to beam
acoustic modes, rescattering, and Langmuir decay instability become
important, the Langdon effect can influence the reflectivity of SRS
by affecting the secondary processes. The
comprehension of Langdon effect on nonlinear evolution and
saturation of SRS  would contribute to a better understanding and
prediction of SRS in inertial confinement fusion.
\end{abstract}
\maketitle

\section{Introduction}
In laser-driven inertial confinement fusion (ICF), stimulated
backward Raman scattering (SRS) is one important laser plasma
instability (LPI), where an incident electromagnetic wave (EMW)
resonantly decays into a backscattered EMW and a forward propagating
electron plasma wave (EPW)~\cite{Montgomery2002TridentLDI}. SRS
needs to be suppressed in ICF, since the backscattered light can
take energy away from the incident laser, and the hot electrons
generated by EPW of SRS can preheat the
capsule~\cite{Lindl2004ICFIndirectIgn}.

Currently, the study of SRS usually assumes a Maxwellian electron
energy distribution
(EEDF)~\cite{Hao2014SRSSBSScatter,Strozzi2008RayBackScatter}.
However, for the laser-irradiated plasma in ICF, the dominant
heating mechanism is inverse bremsstrahlung (IB) heating. The
Langdon effect~\cite{Langdon1980BremssEDF,Matte1988NonMax}, i.e.,
the EEDF tends towards a super-Gaussian form when the IB heating
rate exceeds the electron thermalization rate, can become quite
important under typical hohlraum conditions. It is found that in
the linear convective regime, the Langdon effect can significantly
enhance the gain of SRS and also leads to a shift in the scattered
wavelength~\cite{Qiu2021GaussSpec}. So it is necessary to take
the Langdon effect into account in the investigation of SRS. In
reality, for the widely spreading range of plasma parameters and
laser intensity for typical ICF, there exists some zones where the
SRS can grow absolutely with high reflectivity over a short
distance~\cite{Forslund1975SBS-SRSAnaly,Michel2010SRSSaturationExperimentAbs,Strozzi2017InterPlayLPIHydro,Rose1994LaserSpotNonLinear},
and the nonlinear effects become important to the evolution of SRS.
Then, it is conceivable that both the absolute thresold and the
nonlinear saturation of SRS can be affected by the Langdon effect.
Nevertheless, the investigation of Langdon effect on the absolute
SRS and its nonlinear behavior, which is quite important for the
global understanding and proper modeling of SRS in ICF, is still
lacking.

In this work, the influence of Langdon effect on the nonlinear
evolution of SRS is investigated for a wide range of plasma
parameters at both high and low-intensity laser drives. In the early
growth stage of SRS, it is found that apart from the kinetic
inflation due to trapped
electrons~\cite{Wang2018ConvectiveToAbsoluteRamanInstability,Vu2002KinInflationSRS,Vu2007InflationSRS},
Langdon effect can broaden the parameter range for absolute
SRS modes, and significantly shorten the saturation time for the
absolute SRS in certain parameter region. Over a long timescale
evolution with the development of secondary instabilities such as
the decay of the electron plasma wave to beam acoustic
modes (BAM)~\cite{Yin2006SRSBAM,Yin2006SRSNonMaxBAM,Strozzi2007RamanEAS},
rescattering of the primary scattered
wave~\cite{Hao2021RamanRescatter,Ji2021SRSRescatter} and Langmuir
decay
instability (LDI)~\cite{Russell1999SRSSaturation,Bezzerides1993SRSSaturationLangmuirTurb,Feng2018AntiLangmuirDecayInstability},
the SRS typically demonstrates a change in dominant saturation
mechanism accompanied by a change (usually drop) in the
reflectivity. Under many circumstances, Langdon effect is
found to be important to these secondary instabilities and thus to
the nonlinear saturation mechanism and reflectivity of SRS in the
late evolution stage.

This paper is organized as follows: In Section~\ref{sec:anay},
a three wave coupling model with consideration of the
Langdon effect is given, as well as some theoretical analysis of SRS
in the presence of Langdon effect. Besides, the physics model of a
Vlasov-Maxwell code (VlaMaxW) is presented to account for both the
Langdon effect and nonlinear kinetic effects. In
Section~\ref{sec:Result}, the influences of super-Gaussian
EEDFs on the nonlinear evolution of SRS are investigated for
different stages. In Section~\ref{sec:conc}, the conclusions and
some discussions are given.

\section{Models with Langdon effect considered}
\label{sec:anay}
In SRS, the matching condition requires~\cite{Montgomery2002TridentLDI}
\begin{equation}
  \begin{aligned}
  & \omega_{0}=\omega_s+\omega_l, \\
  & k_{0}=k_s+k_l,
  \end{aligned}
  \label{eq:phasematch}
\end{equation}
where  $\omega_i$ and $k_i$ are the frequencies and wavenumbers with
the subscripts $i=0,\rm s,l$ for the pump wave, the scattered wave
and the EPW, respectively. When the three waves of SRS are dominated
by coherent single modes, the temporal-spatial evolution of SRS can
be described by the coupling equations for their slowly varying
envelopes. Writing the transverse electric fields of the pump wave
and the scattered wave as $\Re[E_0e^{-j\omega_{0c} t+jk_{0c} x}]$,
$\Re[E_se^{-j\omega_{sc} t-jk_{sc} x}]$, and the density
perturbation of the EPW as $\Re[\delta n_l
e^{-j\omega_{lc}t+jk_{lc}x}]$, the envelope equations for the pump
wave and the scattered wave can be written
as~\cite{Strozzi2008RayBackScatter,QiuJ2021SBSSPmode}
\begin{align}
  & (\partial_t+{v}_{0}\partial_x)E_{0}=-\frac{j \omega_{\rm pe}^2}{4\omega_{sc}}\frac{\delta n_l}{n_{e0}}E_s  \label{eq:waveEqnA0},\\
  & (\partial_t-{v}_s\partial_x)E_s=-\frac{j \omega_{\rm pe}^2}{4\omega_{0c}}\frac{\delta n^*_l}{n_{e0}}E_0,
  \label{eq:waveEqnAs}
\end{align}
where $\omega_{\rm pe}$ is the plasma frequency, $n_{e0}$ is the background electron density,
$\omega_{i\rm c}$ and $k_{i\rm c}$ are the fundamental frequency and wavenumber of the pump wave ($i=0$) or the scattered wave ($i=\rm s$) respectively, and
$v_i=c^2k_{i\rm c}/\omega_{i\rm c}$ is the group velocity. Here, $c$ is the light speed.
The fundamental modes satisfy the dispersion relation of EMW
\begin{equation}
  \omega_{i\rm c}^2=\omega_{\rm pe}^2+k_{i\rm c}^2 c^2.
  \label{eq:dispEMW}
\end{equation}
To account for a non-Maxwellian EEDF, a kinetic description of EPW driven ponderomotively can be derived as \cite{Drake1974ParaInstabEM},
\begin{equation}
  (1+\chi_e) \frac{\delta n_l}{n_{e0}}=-\frac{\chi_ee^2{k}_{lc}^2E_{0}E_{s}^*}{2m_e^2 \omega_{\rm pe}^2\omega_{0c}\omega_{sc}},
  \label{eq:eqn2}
\end{equation}
which can be rewritten as
\begin{equation}
  \mathscr{D}\frac{\delta n_l}{n_{e0}}=\frac{e^2{k}_{lc}^2E_{0}E_{s}^*}{2m_e^2 \omega_{\rm pe}^2\omega_{0c}\omega_{sc}},
  \label{eq:eqn2D}
\end{equation}
where $e$ is the electron charge, $m_e$ is the electron mass, and $\mathscr{D}$ is a function of $\omega_l$ and $k_l$
\begin{equation}
  \mathscr{D}(\omega_l,k_l)\equiv -(1+\frac{1}{\chi_e})
  \label{eq:defineD}
\end{equation}
describing the ponderomotive response.
The electron susceptibility $\chi_e$ is given by
\begin{equation}
  \chi_{\rm e}(\omega,k)={\frac{\omega_{\rm pe}^{2}}{k^2\int f_{\rm e0}d\mathbf{v}}}\int \frac{\mathbf{k}\cdot{\partial f_{\rm e0}}/{\partial \mathbf{v}}}{\omega-\mathbf{k}\cdot
  \mathbf{v}}\mathrm{d}\mathbf{v},
\label{eq:xe1d}
\end{equation}
where $f_{e0}$ is the background EEDF.
When Langdon effect is significant, the EEDF has a super-Gaussian form~\cite{Matte1988NonMax,Qiu2021GaussSpec},
\begin{equation}
  f_{\rm e0}(v)=\frac{n_{\rm e}m}{4\pi v_{\rm the}^3\beta_{m}^3\Gamma(3/m)}\exp[-(\frac{v}{\beta_m v_{\textrm{the}}})^{m}],
  \label{eq:fGauss}
\end{equation}
where $m$ is the super-Gaussian exponent, $\Gamma$ is the Gamma
function, $\beta_{m}=\sqrt{3\Gamma(3/m)/\Gamma(5/m)}$, and $v_{\rm
the}=\sqrt{T_{\rm e}/m_{\rm e}}$ is the electron thermal velocity.
Replacing $\mathscr{D}(\omega_l,k_l)$ by the operator
$\mathscr{D}(\omega_{lc}+j\partial_t,k_{lc}-j\partial_x)$, and
taking the slow envelope variation approximation
\cite{Strozzi2008RayBackScatter}
\begin{equation}
  \mathscr{D}(\omega_{lc}+j\partial_t,k_{lc}-j\partial_x)
  \approx  j(\partial\mathscr{D}_r/\partial \omega_l)(\partial_t+v_l \partial_x)+\mathscr{D}(\omega_{lc},k_{lc}),
  \label{eq:epsoperator}
\end{equation}
the envelope equation of EPW can be derived as
\begin{equation}
  (\partial_t+v_l\partial_x+\nu_{l}+j\delta\omega_l)
  \frac{\delta n_l}{n_{e0}}=-\frac{j}{\partial \mathscr{D}_r/\partial \omega_l} \frac{e^2{k}_{lc}^2E_{0}E_{s}^*}{2m_e^2 \omega_{\rm pe}^2\omega_{0c}\omega_{sc}}.
  \label{eq:eqnEPWEnv}
\end{equation}
Here
only $\mathscr{D}_r\equiv \mathrm{Re}[\mathscr{D}]$ is retained in the derivatives for
$|\Im[\partial\mathscr{D}/\partial \omega_l]|\ll |\partial\mathscr{D}_r/\partial \omega_l|$. The group velocity of the EPW
\begin{equation}
v_l=-\frac{\partial \mathscr{D}_r/\partial k_l}{\partial \mathscr{D}_r/\partial \omega_l},
  \label{eq:gvEPW}
\end{equation}
the Landau damping of EPW
\begin{equation}
  \nu_l=\frac{\Im[\mathscr{D}]}{\partial \mathscr{D}_r/\partial \omega_l},
  \label{eq:nulD}
\end{equation}
and the frequency mismatch
\begin{equation}
  \delta\omega_l=-\frac{\mathscr{D}_r}{\partial \mathscr{D}_r/\partial \omega_l}.
  \label{eq:OmglD}
\end{equation}
Notice that in this approach, the Langdon effect can be easily
incorporated just by using Eq.~(\ref{eq:fGauss}) when calculating
$\chi_e$. Furthermore, the ponderomotive operator $\mathscr{D}$
instead of the dielectric operator $\epsilon=1+\chi_e$
\cite{Strozzi2008RayBackScatter} is adopted to describe the response
of the EPW field to the ponderomotive drive from the beating of the
pump wave and the scattered wave. Especially for large
$k_l\lambda_{\rm De}$ with $\lambda_{\rm De}=v_{\rm the}/\omega_{\rm
pe}$ being the Debye length, this modification is necessary since
$\Im[\partial\mathscr{D}/\partial \omega_l]\ll
|\partial\mathscr{D}_r/\partial \omega_l|$ can be satisfied for
$k_l\lambda_{\rm De}$ ranging from 0 to 1  while the assumption
$\Im[\partial\epsilon/\partial \omega_l]\ll
|\partial\epsilon_r/\partial \omega_l|$ is broken when
$k_l\lambda_{\rm De}$ is near $0.4$.

It is convenient to renormalize the wave amplitudes as
\begin{equation}
  \begin{aligned}
    a_0 &\equiv \frac{E_0}{|E_{\rm 0L}|},\\
    a_s &\equiv \frac{E_s}{|E_{\rm 0L}|}\sqrt{\frac{\omega_{0c}}{\omega_{sc}}},\\
    a_l &\equiv \frac{E_l}{|E_{\rm 0L}|}\sqrt{\frac{\omega_{0c}}{2}\frac{\partial \mathscr{D}_r}{\partial \omega_l}}\approx \frac{E_l}{|E_{\rm 0L}|}\frac{\sqrt{\omega_{0c}\omega_{lc}}}{\omega_{\rm pe}},
  \end{aligned}
  \label{eq:renormfac}
\end{equation}
where $E_l=je\delta n_l/\epsilon_0 k_l$ is the electrostatic field
of EPW, and $E_{\rm 0L}$ is the pump field amplitude incident at the
left boundary. Then, the three wave coupling (TWC) equations
of SRS can be recast into the following simplified form
\begin{align}
  (\partial_t+{v}_{0}\partial_x)a_0&=-\gamma_0 a_sa_l,
  \label{eq:wave3_a0} \\
  (\partial_t-{v}_s\partial_x)a_s&=\gamma_0 a_0a_l^*,
  \label{eq:wave3_as} \\
  (\partial_t+v_l\partial_x+\nu_{l}+j\delta\omega_l)a_l&=\gamma_0 a_0a_s^*.
  \label{eq:wave3_al}
\end{align}
where $\gamma_0$ is the homogeneous growth rate of SRS when the
Landau damping and frequency mismatch are ignored. By substituting
the assumed solution forms $E_{s}, \delta n_l/n_0\propto e^{\gamma_0
t}$ into Eq.~(\ref{eq:waveEqnAs}) and Eq.~(\ref{eq:eqnEPWEnv})
\begin{equation}
  \begin{aligned}
    \gamma_0=\frac{k_lv_{\rm os}}{2\sqrt{2\omega_s(\partial \mathscr{D}_r/\partial \omega_l)}} \approx
    \frac{k_lv_{\rm os}}{4}\frac{\omega_{\rm pe}}{\sqrt{\omega_l\omega_s}}
  \end{aligned}
  \label{eq:growrate}
\end{equation}
can be obtained, where $v_{\rm os}=e|E_0|/m_e\omega_{0c}$ is the electron quiver velocity. 
Then, the instantaneous reflectivity $R$ can be determined from
the scattered wave amplitude emergent from the left boundary
\begin{equation}
  R\equiv \frac{|E_{\rm sL}|^2v_s/\omega_{sc}}{|E_{\rm 0L}|^2v_0/\omega_{0c}}=|a_{\rm sL}|^2\frac{v_s}{v_0}.
  \label{eq:Reflect}
\end{equation}
In this definition, $R<1$ limited by pump depletion can be
guaranteed if $R$ eventually reaches a constant
value~\cite{Forslund1975SBS-SRSAnaly}.

SRS can be in the convective instability regime or in the absolute
instability
regime~\cite{Forslund1975SBS-SRSAnaly,Steinberg1986AbsConvComptonSRS,Hall1968InstaAbsConv}.
The convective instability typically occurs under strong damping
condition. The convective gain coefficient $\kappa_{\rm R}$ can be
derived by assuming the solution form $a_{s}, a_l\propto
e^{-\kappa_{\rm R}x}$, where the temporal and spatial derivatives in
Eq.~(\ref{eq:wave3_al}) can be ignored compared to the Landau
damping, yielding
\begin{equation}
  \kappa_{\rm R}=\frac{\gamma_0^2}{v_s}\Re[\frac{1}{\nu_l+j\delta \omega_l}]
  =\frac{k_l^2v_{\rm os}^2}{8c^2k_s}\Im[\frac{\chi_e}{1+\chi_e}]
\end{equation}
which is just the classical formula for the kinetic convective gain
coefficient of SRS~\cite{Drake1974ParaInstabEM,Qiu2021GaussSpec}.
The strong damping condition $\nu_l\gg \kappa_Rv_l$ such implies
$\nu_l\gg \gamma_0\sqrt{v_l/v_s}$, under which the saturated
reflectivity with pump depletion considered can be analytically
determined from the TWC model, given by the Tang's
formula~\cite{Tang1966SatSpecSBS,Forslund1975SBS-SRSAnaly},
\begin{equation}
  R(1-R)=\varepsilon \{\exp[G_{\rm R}(1-R)]-R\},
  \label{eq:Tang}
\end{equation}
where $\varepsilon=|a_{\rm s,Right}|^2v_s/v_0$ is determined by the seed light intensity at the right boundary,
and $G_{\rm R}=2\kappa_{\rm R}L$ is the energy gain of SRS for an amplification length $L$.

The absolute instability usually occurs under strong laser drive or weak Landau damping.
It is of important concern for the SRS control in ICFs~\cite{Michel2010SRSSaturationExperimentAbs},
since SRS keeps growing until saturated by nonlinear effects,
generally leading to a large reflectivity.
When nonlinear effects can be ignored,
by analyzing Eqs.~(\ref{eq:wave3_as}-\ref{eq:wave3_al}) with Laplace transform,
the absolute instability condition can be derived as~\cite{Forslund1975SBS-SRSAnaly,Bezzerides1996SBSTwoIon}
\begin{equation}
  2\gamma_0\sqrt{v_l/v_s}\geq\nu_l,
  \label{eq:abscond}
\end{equation}
and the absolute growth rate is
\begin{equation}
  \gamma_{\rm abs}=2\gamma_0\sqrt{v_l/v_s}-\nu_l.
  \label{eq:absrate}
\end{equation}

This TWC model can describe the evolution of SRS both in the
convective and absolute regimes with the consideration of Langdon
effect and the the pump depletion, but excludes all the nonlinear
kinetic effects such as reduction of Landau damping and frequency
shift due to trapped electrons. So as simple as it is, the TWC model
is helpful for understanding the initial growth of SRS and identify
the onset of nonlinear kinetic effects.

As a super-Gaussian EEDF can significantly reduce $\nu_l$~\cite{Qiu2021GaussSpec}, it is expected
that the absolute instability condition is easier to be met when Langdon effect is considered.
Nevertheless, in most cases the linear condition fails to be a good criterion to judge whether absolute growth can occur,
since the kinetic inflation~\cite{Vu2007InflationSRS,Wang2018ConvectiveToAbsoluteRamanInstability,Ellis2012SRSConvective},
wherein the convective SRS is transformed into absolute SRS due to electron trapping effects,
is found to be quite important for SRS. To take into account the nonlinear kinetic effects,
an one spatial dimensional and one velocity dimensional (1D1V) Vlasov-Maxwell code (VlaMaxW) has been developed, which solves the Vlasov equations
\begin{align}
  &\frac{\partial f_\alpha}{\partial t}+\frac{\partial}{\partial x}[\frac{p_{xj}}{m_\alpha}f_\alpha]+\frac{\partial }{\partial p_{x\alpha}}[V_{px\alpha}f]=0, \\
  & V_{px\alpha}\equiv q_\alpha E_x-\frac{q_\alpha^2}{2m_\alpha}\frac{\partial A_y^2}{\partial x},
\end{align}
together with Maxwell equations
\begin{align}
  &\frac{\partial B_z}{\partial t}=-\frac{\partial E_y}{\partial x} \\
  &\frac{\partial E_y}{\partial t}=-\frac{\partial (c^2B_z)}{\partial x}-\frac{J_y}{\epsilon_0} \\
  &\frac{\partial E_x}{\partial t}=-\frac{J_x}{\epsilon_0} \\
  & E_y=-\frac{\partial A_y}{\partial t} \\
  & J_{y}=-\sum_\alpha\frac{q_\alpha^2}{m_\alpha}\int f_\alpha dp_{x\alpha}A_y \\
  & J_x=\frac{q_\alpha}{m_\alpha}\int f_\alpha dp_{x\alpha}p_{x\alpha}
\end{align}
by numerical methods similar to Ref.
\cite{Strozzi2017InterPlayLPIHydro}. Here the cold plasma
approximation for the transverse motion $p_{y\alpha}=-q_\alpha A_y$
is assumed, and subsript $\alpha=e, i$ for electron and ion,
respectively. $m_\alpha$ is the mass, $q_\alpha$ is the charge, $\epsilon_0$ is
the vacuum permittivity, $p_{x\alpha}$ is the momentum along $x$ direction,
$A_y$ is the vector potential, $E_x$ is the electrostatic field,
$E_y$ and $B_z$ are the transverse electric and magnetic fields,
respectively.

In following, the Vlasov-Maxwell simulation is primarily used to
investigate the influence of super-Gaussian EEDF on the nonlinear
evolution (e.g., inflation and saturation) of SRS, while the TWC
simulation is conducted to provide a valuable reference to help
comprehend the simulation results, where the fundamental mode of the
EPW is chosen at the peak value of $\kappa_{\rm R}$. For small
$k_l\lambda_{\rm De}$, it is quite close to the natural mode of EPW
which satisfies $1+\chi_e(\omega_{l},k_{l})=0$, and also $\delta
\omega_l\approx 0$. While for $k_l\lambda_{\rm De}>0.3$, this can
deviate significantly from the natural mode, and also $\delta
\omega_l$ can be nonzero.

\section{Langdon effect on the nonlinear evolution of SRS}
\label{sec:Result}

\begin{figure*}[!hbtp]
  \centering
  \includegraphics[angle=0,width=0.97\textwidth]{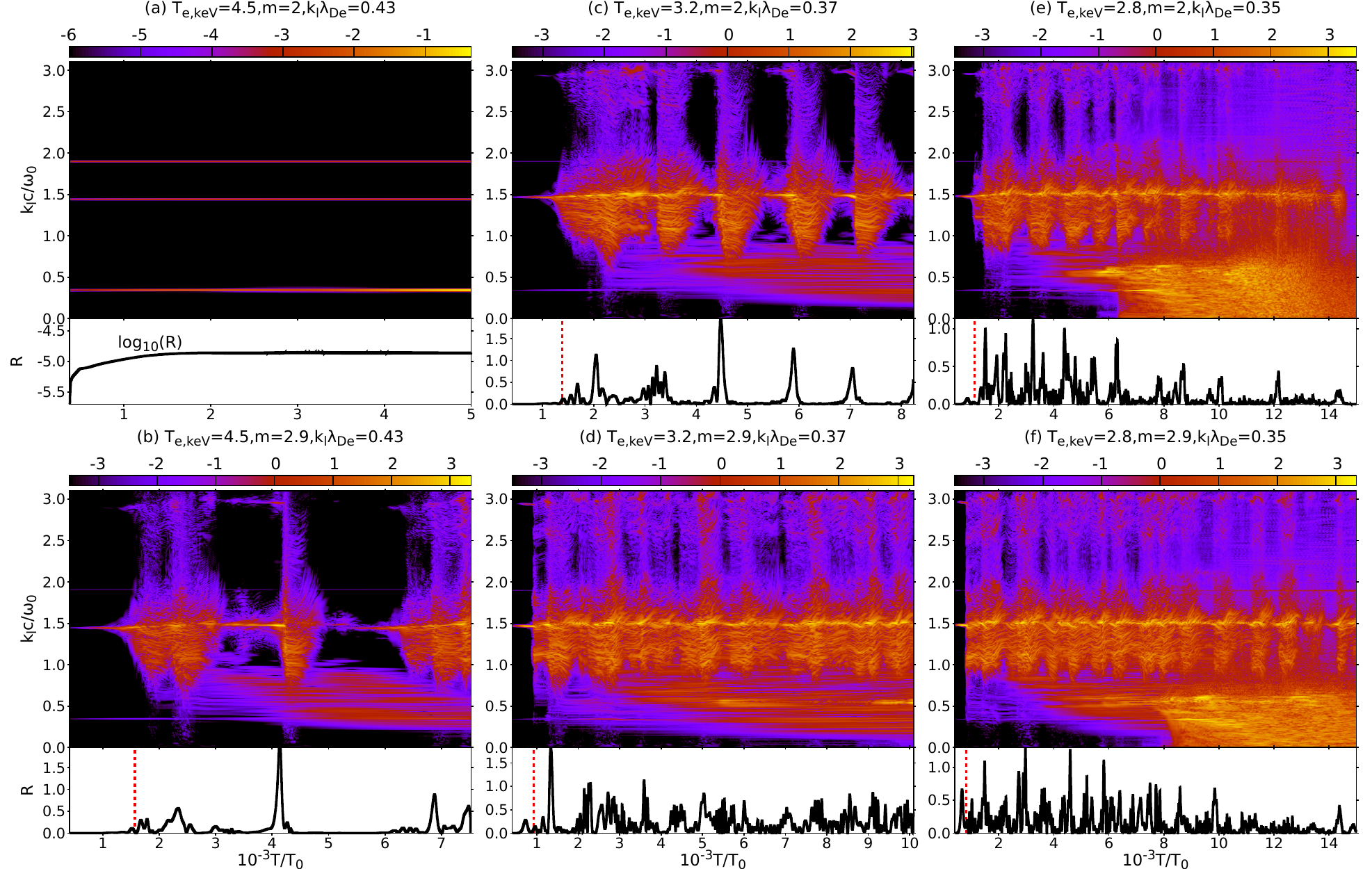}
  \caption{
    The evolution of $k_l$-spectrum of the EPW field [$\rm log_{10}|E_l(k_l)|^2$] for several cases with $T_e=4.5, 3.2, 2.8$keV and $m=2, 2.9$ when $I_{15}=2.82$.
    The corresponding reflectivity versus time is displayed in the bottom panels, where dividing-line between the early stage and the later stage is marked by the red dotted vertical lines.
The condition $n_e=0.1~n_c$, $\lambda_0=351~\rm nm$ for a homogeneous He plasma with $T_i=T_e/5$ and length of 200$\lambda_0$ is taken.
  }
  \label{Fig:KSpectrumHkLD}
\end{figure*}

\begin{figure*}[!hbtp]
  \centering
  \includegraphics[angle=0,width=0.98\textwidth]{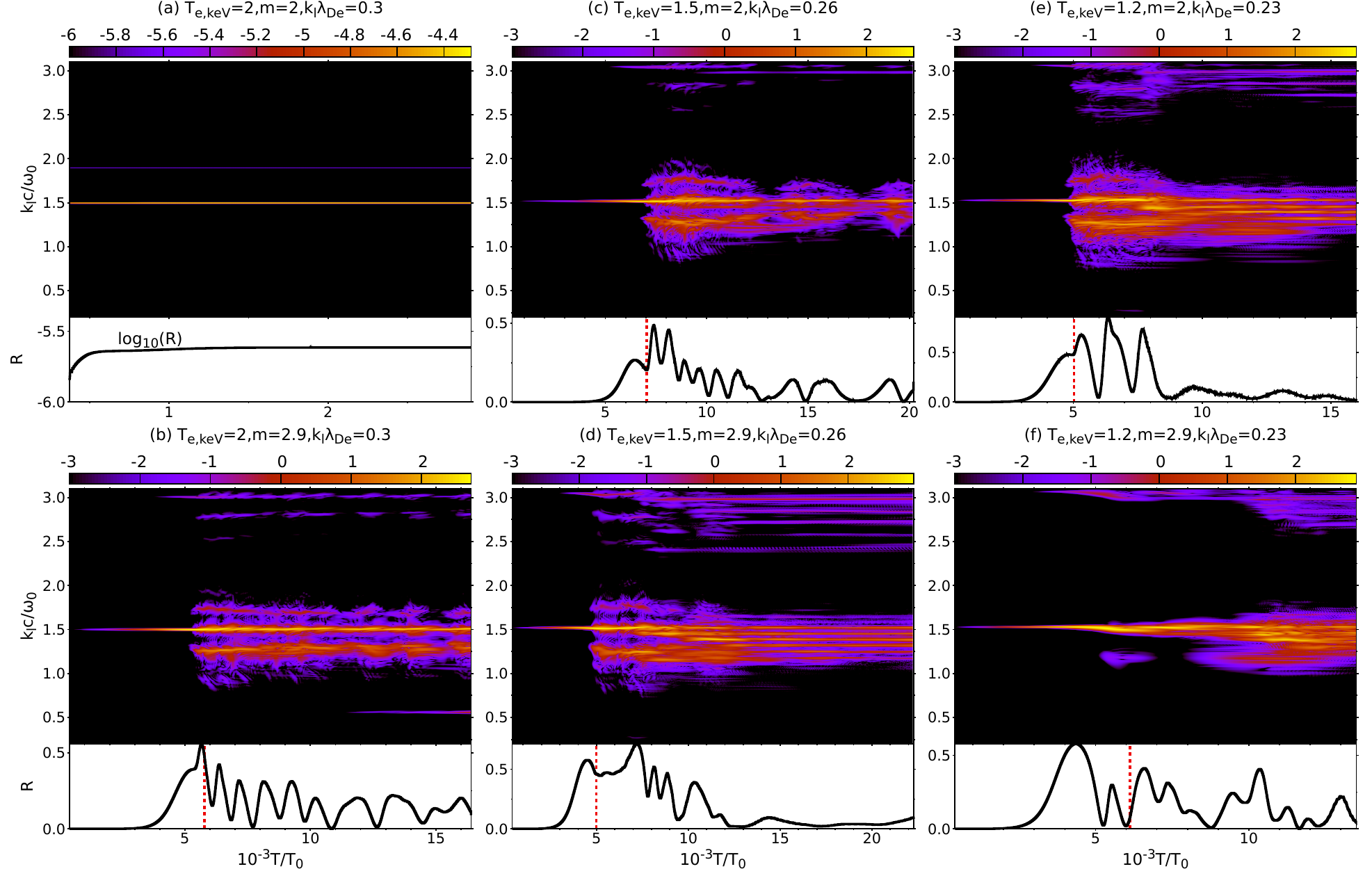}
  \caption{
    The evolution of $k_l$-spectrum of the electrostatic field [$\rm log_{10}|E_z(k_l)|^2$] for several cases with $T_e=2, 1.5, 1.2$keV and $m=2, 2.9$ when $I_{15}=0.11$.
    The corresponding reflectivity versus time is displayed in the bottom panels, where dividing-line between the early stage and the later stage is marked by the red dotted vertical lines.
The condition $n_e=0.1~n_c$, $\lambda_0=351~\rm nm$ for a homogeneous He plasma with $T_i=T_e/5$ and length of 200$\lambda_0$ is taken.
  }
  \label{Fig:KSpectrumLkLD}
\end{figure*}

In the Vlasov-Maxwell and TWC model simulations, a homogeneous He
plasma with length $L_0=200\lambda_0$ is assumed, where the laser
vacuum length $\lambda_0 = 0.351~\rm \mu m$, $L_0$ is on the speckle
length scale $\sim 8f^2\lambda_0$ of a realistic laser beam with $f$
being the f/number~\cite{Lindl2004ICFIndirectIgn}, and the low-Z He
plasma is the typical environment for significant SRS generation in
ICFs~\cite{Lindl2004ICFIndirectIgn,Hall2017GasFillNIF}. In the
Vlasov-Maxwell simulation, also two additional collision layers
with lengths $2 \times 20\lambda_0$ and ramp electron density
profiles are appended to both sides of the plasma to eliminate
effects of the sheath field. One linearly polarized laser beam with
intensity $I_0$ is incident from the left boundary at $t=0$ with a
rise time $20T_0$, while the seed light with frequency $\omega_{sc}$
and intensity $I_s=10^{-6} I_0$ is incident from the right boundary
at $t=100T_0$ with a rise time $20 T_0$, where $T_0=2\pi/\omega_0$
is the laser period. Since the typical range of super-Gaussian
exponent $m$ for low-Z He plasma is between $2$ and $3$ as analyzed
in Ref.~\cite{Qiu2021GaussSpec}, simulation cases with $m=2$ and
$m=2.9$ are compared to demonstrate impacts of Langdon effect in
this work. As known, $k_l\lambda_{\rm De}\propto
\sqrt{T_e/n_e}$, is a key parameter to determine the saturation
mechanism of
SRS~\cite{Kline2006LangmuirWaveNonLinearRegime,Montgomery2002TridentLDI},
so we widely scan $k_l\lambda_{\rm De}$ from $0.48$ to $0.19$ in
simulations, by changing $T_e$ from $5$~keV to $0.8$~keV with
$n_e=$0.1$~n_c$ ($n_c$ is the critical density). Laser intensity is
typically chosen as $I_{15}=2.82$ ($I_{15}=I_0/10^{15}~\rm
Wcm^{-2}$), which is achievable in small laser speckles in ICF.
Besides, a low laser intensity $I_{15}=0.11$ is also used for the
low $k_l\lambda_{\rm De}$ cases.

In the Vlasov-Maxwell simulations, the temporal evolution of the
electrostatic field in the $k_l$-space together with the
instantaneous reflectivity are shown with different $T_e$ and $m$ in
Fig.~\ref{Fig:KSpectrumHkLD} and Fig.~\ref{Fig:KSpectrumLkLD} for
$I_{15}=2.82$ and $I_{15}=0.11$, respectively. Except the cases
shown in Fig.~\ref{Fig:KSpectrumHkLD}(a) and
Fig.~\ref{Fig:KSpectrumLkLD}(a), where the SRS reflectivity is
saturated at a low level due to convective
saturation~\cite{Forslund1975SBS-SRSAnaly,Bezzerides1996SBSTwoIon};
the nonlinear effects are obvious in other cases, where the onset of
early-stage saturation of the reflectivity generally occurs before
significant broadening of the $k_l$-spectrum induced by the
secondary instabilities and even the cascaded instabilities.
Therefore, the SRS evolution can be approximately divided into the
early-stage where the secondary instabilities play negligible roles, and
the late-stage where the secondary instabilities dominate and result
in non-stationary variation of the reflectivity. In the following
Subsection~\ref{sec:earlystage}, the influences of Langdon effect on
the early growth and saturation of SRS are mainly studied. Then, the
differences in the dominant saturation mechanism and reflectivity of
SRS for different $m$ in the late-stage are discussed in
Subsection~\ref{sec:latestage}.

\subsection{The early-stage growth and saturation of SRS}
\label{sec:earlystage}

\begin{figure*}[!hbtp]
  \centering
  \includegraphics[angle=0,width=0.95\textwidth]{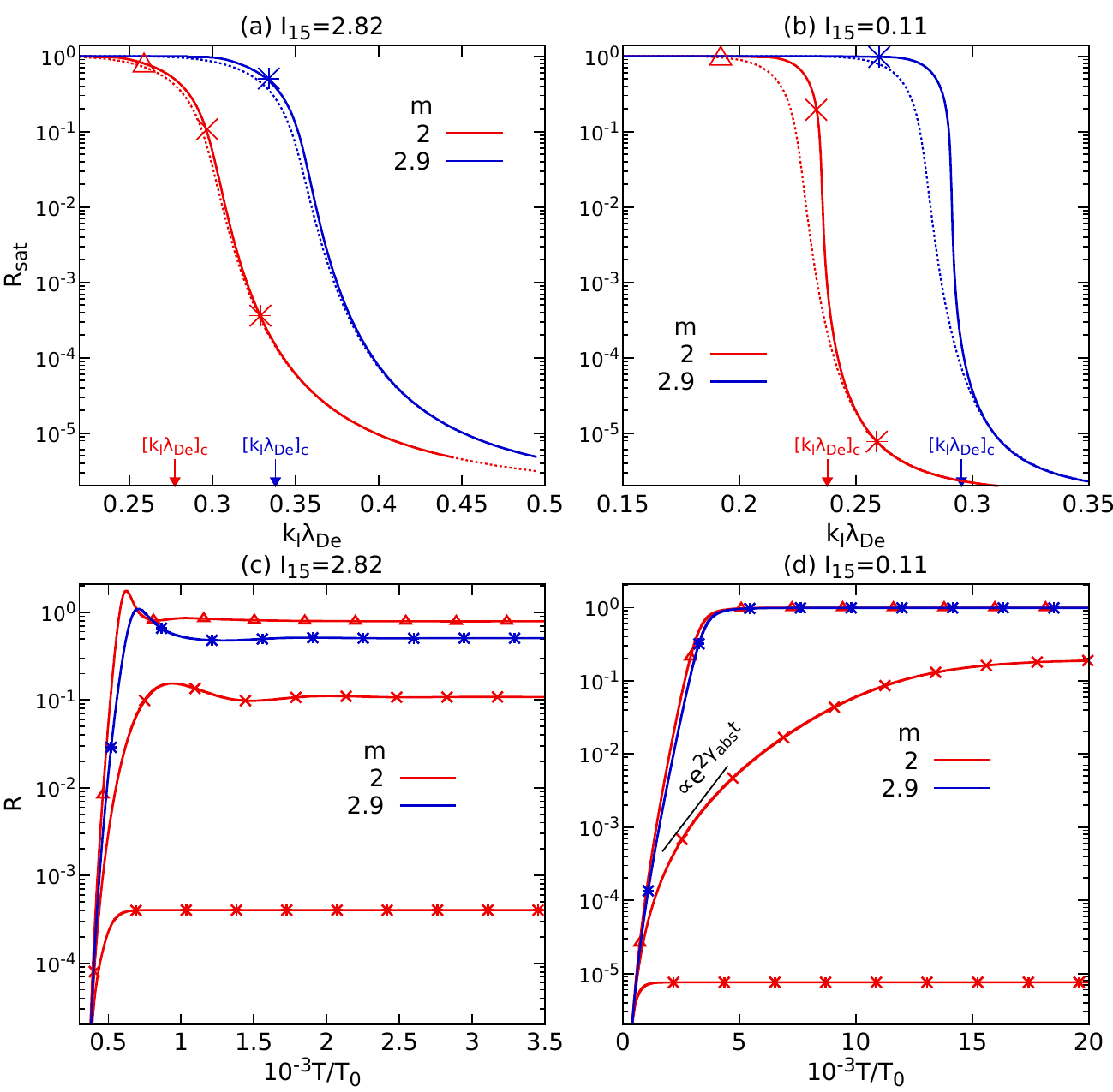}
  \caption{
    The saturated reflectivity versus $k_l\lambda_{\rm De}$ from the TWC model (solid lines) for (a) $I_{15}=2.82$ and (b) $I_{15}=0.11$ at $m=2$ (in red) and $m=2.9$ (in blue).
    Correspondingly, the temporal evolution of the reflectivity for cases indicated by asterisks, crosses and triangles in (a) and (b)
    is presented in (c) and (d), respectively.
    In (a) and (b), the reflectivity versus $k_l\lambda_{\rm De}$ calculated by the Tang's formula is also plotted as dotted lines,
    while $[k_l\lambda_{\rm De}]_c$ for absolute growth
    calculated from the linear criterion (\ref{eq:abscond})
    is indicated for both $m=2$ and $m=2.9$.
  }
  \label{Fig:TWCRef}
\end{figure*}

As shown in Fig.~\ref{Fig:KSpectrumHkLD} and
Fig.~\ref{Fig:KSpectrumLkLD}, SRS grows convectively with a low
saturation level at the larger $k_l\lambda_{\rm De}$, but grows
quickly to a high early-stage saturation level at the smaller
$k_l\lambda_{\rm De}$. This issue can be predicted by the linear
criterion  $\nu_{l0}\leq 2\gamma_0\sqrt{v_l/v_s}$ for absolute SRS,
where $\nu_{l0}$ is the initial Landau damping.
Since $\nu_{l0}$ is a rapidly decreasing function of
$k_l\lambda_{\rm De}$, there exists one critical $[k_l\lambda_{\rm
De}]_c$ at which the equality of the criterion can be satisfied. For
$k_l\lambda_{\rm De}>[k_l\lambda_{\rm De}]_{\rm c}$, the SRS growth
is convective. The reflectivity can saturate at a lower level and
the nonlinear effects are quite weak, as presented in
Fig.~\ref{Fig:TWCRef}(c-d). In such cases, the reflectivity
calculated by the Tang's formula (\ref{eq:Tang}) agrees well with
the simulation results of the TWC model, as shown in
Fig.~\ref{Fig:TWCRef}(a-b). For $k_l\lambda_{\rm
De}<[k_l\lambda_{\rm De}]_{\rm c}$, the SRS growth is absolute,
implying that SRS can keep growing until saturated by nonlinear
effects, as shown by the time evolution of reflectivity in
Fig.~\ref{Fig:TWCRef}(c-d), which is calculated by the TWC model and
hence the only nonlinear effect is the pump depletion. Since
nonlinear effects that become important only at large reflectivity
are necessary to prevent further growth of SRS, the reflectivity for
absolute SRS growth can not be too small. Consequently, when
$[G_{\rm R}]_{\rm c}$ at $[k_l\lambda_{\rm De}]_{\rm c}$ is small
and hence the convectively saturated level is low, there would be a
sharp increase of reflectivity (especially, much sharper than the
prediction by the Tang's model) at the transition from convective to
absolute SRS, as exampled in Fig.~\ref{Fig:TWCRef}(b). In
comparison, the change of $R$ near $[k_l\lambda_{\rm De}]_{\rm c}$
is much more gradual in Fig.~\ref{Fig:TWCRef}(a), since for $[G_{\rm
R}]_{\rm c}>15$ (cf. Table~\ref{Tab:absthr}), the reflectivity due
to convective amplification is already sufficiently large to incur
the nonlinear saturation effects.

In addition to decreasing $k_l\lambda_{\rm De}$, increasing $m$ also
leads to a decrease of $\nu_{l0}$, and thus can result in the
transition from convective to absolute SRS in the certain
region of $k_l\lambda_{\rm De}$. Comparing the lines with asterisks
in Fig.~\ref{Fig:TWCRef}(c) and (d), for the same laser and plasma
parameters, SRS is in the convective regime when $m=2$ but can grow
absolutely and eventually saturated by nonlinear effects when
$m=2.9$. Consequently, $[k_l\lambda_{\rm De}]_{\rm c}$ becomes
larger for a greater $m$, as listed in Table~\ref{Tab:absthr},
indicating that the Langdon effect can broaden the parameter range
for absolute SRS instability as shown in Fig.~\ref{Fig:TWCRef}. In
Table~\ref{Tab:absthr}, it is found that the difference in
$[k_l\lambda_{\rm De}]_{\rm c}$ for different $m$ results in close
(slightly higher for increasing $m$) values of $[\nu_{l0}]_{\rm c}$
in the linear prediction. This is because in Eq.~(\ref{eq:abscond}),
$\gamma_0$ and $v_s$ are almost independent of $k_l\lambda_{\rm
De}$, while $v_l\approx 3v_{\rm the}^2k_l/\omega_l\propto
(k_l\lambda_{\rm De})^2$ increases weakly with larger
$[k_l\lambda_{\rm De}]_c$ at greater $m$, in contrast to the strong
rise of $\nu_{l0}$ with decreasing $k_l\lambda_{\rm De}$.

\begin{table}[!ht]
\centering
\begin{threeparttable}
\caption{Summary of critical parameters for the transition from
convective to absolute SRS growth from the linear TWC model and from
the Vlasov-Maxwell simulation. The homogeneous He plasma with
$n_e=0.1~n_c$, $T_i=T_e/5$ and length of $200\lambda_0$
($\lambda_0=351~\rm nm$) is taken. } \label{Tab:absthr}
\begin{tabular}{cccccccc}
\toprule
\multirow{2}{*}{$I_{15}$} & \multirow{2}{*}{$m$} &  \multicolumn{3}{c}{Linear TWC} & \multicolumn{3}{c}{Vlasov-Maxwell}  \\
&  &  $[k_l\lambda_{\rm De}]_{\rm c}$ & $\frac{[\nu_{l0}]_{\rm c}}{0.01\omega_0}$ & $[G_{\rm R}]_{\rm c}$ & $[k_l\lambda_{\rm De}]_{\rm c}$ & $\frac{[\nu_{l0}]_{\rm c}}{0.01\omega_0}$ & $[G_{\rm R}]_{\rm c}$  \\
\midrule
\multirow{2}{*}{2.82} & 2 & 0.278 & 0.208 & 20.5 & 0.42 & 2.58 & 1.68  \\
 & 2.9 & 0.338 & 0.241 & 16.4 & 0.465 & 2.15 & 1.84  \\
 \multirow{2}{*}{0.11} & 2 & 0.238 & 0.0358 & 4.99 & 0.269 & 0.154 & 1.12 \\
 & 2.9 & 0.295 & 0.0429 & 3.92 & 0.316 & 0.112 & 1.45 \\
\bottomrule
\end{tabular}
\end{threeparttable}
\end{table}

\begin{figure*}[!hbtp]
  \centering
  \includegraphics[angle=0,width=0.95\textwidth]{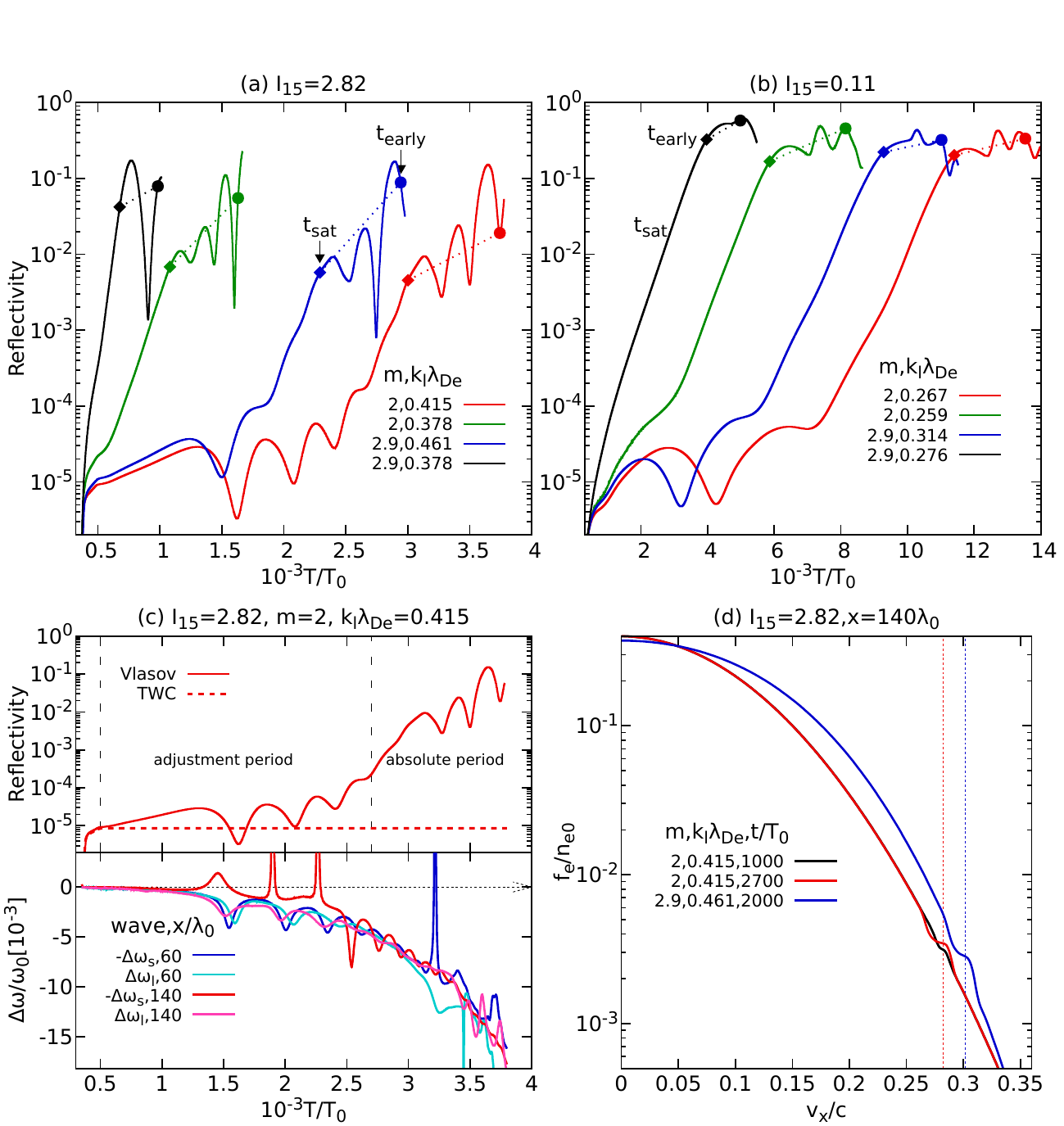}
  \caption{
    The early-stage evolution of the instantaneous reflectivity for (a) $I_{15}=2.82$ and (b) $I_{15}=0.11$.
    The circles denote $t_{\rm early}$ defined as the time until when 90\% of the EPW energy is contained within $[0.98k_{lc},1.02k_{lc}]$.
    The diamonds denote $t_{\rm sat}$ corresponding to the turning point of the reflectivity, which is determined  by fitting $\log_{10}R$ versus $t$ with a piecewise linear function.
    The case $m=2$ and $k_l\lambda_{\rm De}=0.415$ in (a) is replotted in (c),
    where the reflectivity versus time is compared to the prediction of TWC model in the upper panel,
    while the frequency shifts of the EPW and the scattered wave are displayed in the bottom panel for distances $x=60\lambda_0$ and $x=140\lambda_0$ from the left boundary.
    In (d), the flatted EEDF is shown for two cases in (a) at $m=2$ and $m=2.9$, where $v_{\rm phl}$ is indicated by vertical dotted lines.
  The condition $n_e=0.1~n_c$, $\lambda_0=351~\rm nm$ for a homogeneous He plasma with $T_i=T_e/5$ and length of 200$\lambda_0$ is taken.
  }
  \label{Fig:SRS_early}
\end{figure*}

Considering the nonlinear kinetic effects can not be
included by the TWC model, the Vlasov-Maxwell simulation results
for some cases 
are shown in
Fig.~\ref{Fig:SRS_early}. Due to the nonlinear kinetic effects, the
convective SRS in the linear prediction can change into 
absolute SRS. One example is presented in
Fig.~\ref{Fig:SRS_early}(c), where the initial growth of SRS agrees
well with the TWC model before $t<500T_0$, after which the TWC model
predicts convective saturation while the Vlasov-Maxwell simulation
exhibits the continual growth of SRS towards much higher
reflectivity. A detailed investigation shows that with the growth of
the EPW field, resonant electrons with $v_x\approx v_{\rm phl}$ are
trapped, resulting in flattening of the EEDF around $v_{\rm phl}$,
as shown in Fig.~\ref{Fig:SRS_early}(d). Correspondingly, the Landau
damping is
reduced~\cite{ONeil1965CollisionlessDamping,Yampolsky2009ModelNLLandau}
while the EPW frequency is downshifted causing a frequency
mismatch~\cite{Morales1972NonLinearFrqShift,Berger2013KinWave,Vu2001ParticleTrappingDetuning}.
This in turn induces a frequency upshift of the scattered wave that
tends to restore the frequency matching resonance, as shown in the
bottom panel of Fig.~\ref{Fig:SRS_early}(c) for the adjustment
period $500T_0<t<2700T_0$. The frequency mismatch induces a phase
mismatch $\delta_{\rm mis}$ and thus a reduction of growth rate by
$\cos\delta_{\rm mis}$, impairing the SRS growth, while the
nonlinear reduction of $\nu_l$ favors the SRS growth. In the case
shown in Fig.~\ref{Fig:SRS_early}(c), the competition between these
two factors results in oscillation of the reflectivity during
$1500T_0<t<2500T_0$, wherein significant frequency mismatch exists in the plasma region further away from the left boundary (e.g. $x/\lambda_0=140$ in Fig.~\ref{Fig:SRS_early}c). 
After a period of adjustment, ultimately at $t>2700T_0$ over a large
plasma region the frequency upshift of the scattered wave becomes
sufficiently large to compensate the frequency downshift of the EPW,
reducing the frequency mismatch to a low level. Consequently, the
nonlinear absolute growth rate $\gamma_{\rm abs,NL}\approx
2\gamma_0\sqrt{v_l/v_s}\cos\delta_{\rm mis}-\nu_{\rm NL}$ exceeds
zero over a large plasma region, and SRS enters into the absolute
growth period. Until when $t>3100T_0$ the EPW amplitude in the
plasma region near the left boundary (e.g. $x/\lambda_0=60$ in
Fig.~\ref{Fig:SRS_early}c) is so large that the frequency downshift
of the EPW can not be completely compensated to maintain the SRS
resonance, resulting in saturation of the absolute SRS growth.

To illustrate the Langdon effect on the early-stage
saturation of SRS in Vlasov-Maxwell simulation, we need to measure
the early-stage saturated reflectivity $R_{\rm sat,e}$.
However, as shown in Fig.~\ref{Fig:KSpectrumHkLD}(b-f) and
Fig.~\ref{Fig:KSpectrumLkLD}(b-f), for the absolute SRS, the
reflectivity versus time after the saturation is irregular and
nonstationary, making it necessary to clearly define $R_{\rm sat,e}$
in a reasonable way. Here we define $R_{\rm sat,e}$ as the averaged
reflectivity over the time period from the onset of early-stage
saturation ($t_{\rm sat}$) until the onset of significant secondary
instabilities ($t_{\rm early}$). Conveniently, $t_{\rm sat}$ can be
identified as the time when the growth of the reflectivity becomes
flattened, and $t_{\rm early}$ can be defined as the time until when
$90\%$ of the EPW field energy is contained within $[0.98k_{\rm
lc},1.02k_{\rm lc}]$, as indicated in Fig.~\ref{Fig:SRS_early}(a-b).
For the convective SRS growth, on the other hand, the reflectivity
would eventually become constant with time or in some cases
oscillate in a regular way and hence have a constant mean value;
correspondingly, it is natural to recognize the saturated
reflectivity $R_{\rm sat}$ as the steady (mean) value of the
reflectivity.

\begin{figure*}[!hbtp]
  \centering
  \includegraphics[angle=0,width=0.95\textwidth]{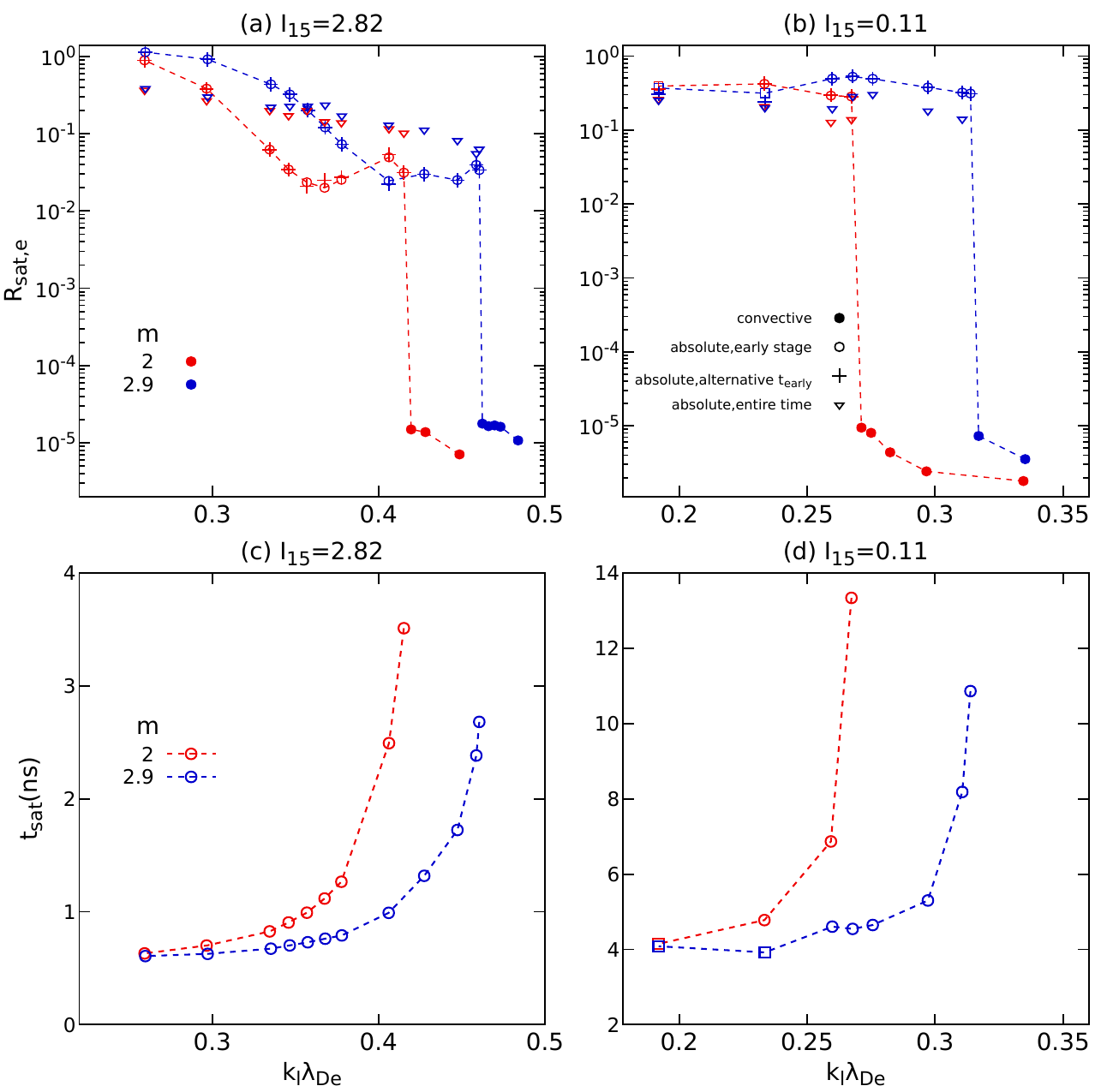}
  \caption{
    (a-b) $R_{\rm sat,e}$ versus $k_l\lambda_{\rm De}$
     at different $m$ in Vlasov-Maxwell simulations for (a) $I_{15}=2.82$ and (b) $I_{15}=0.11$.
     Three types of cases are distinguished: (I) absolute growth when the $k_l$-spectrum broadening of EPW due to secondary instabilities is abrupt (open circles).
    (II) absolute growth when the $k_l$-spectrum broadening of EPW is gradual (open squares).
    (III) convective growth with rather weak nonlinear effects (solid circles).
    For comparison,
    $R_{\rm sat,e}$ calculated using an alternative definition of $t_{\rm early}$ (90\% of the EPW field energy is contained within $[0.95k_{lc},1.05k_{lc}]$) is displayed by the plus symbols,
    while the time-averaged reflectivity over the entire simulation time with $t>t_{\rm sat}$ is shown as open triangles.
    (c-d) $t_{\rm sat}$ versus $k_l\lambda_{\rm De}$ at different $m$
    for cases with absolute growth.
  The condition $n_e=0.1~n_c$, $\lambda_0=351~\rm nm$ for a homogeneous He plasma with $T_i=T_e/5$ and length of 200$\lambda_0$ is taken.
  }
  \label{Fig:SRSSatLevel_kLD}
\end{figure*}

In Vlasov-Maxwell simulation, $R_{\rm sat,e}$ versus
$k_l\lambda_{\rm De}$ is shown in
Fig.~\ref{Fig:SRSSatLevel_kLD}(a-b) for different $m$, whereas
$t_{\rm sat}$ versus $k_l\lambda_{\rm De}$ is displayed in
Fig.~\ref{Fig:SRSSatLevel_kLD}(c-d) for cases with absolute SRS
growth. As a comparison, $R_{\rm sat,e}$ with an alternative
definition of $t_{\rm early}$, i.e. the time until when 90\% of the
EPW energy is contained within $[0.95k_{lc},1.05k_{lc}]$, as well as
the time-averaged reflectivity over the entire simulation time with
$t\geq t_{\rm sat}$, is also shown. As seen, $R_{\rm sat,e}$ is
insensitive to the individual choice in the definition of $t_{\rm
early}$. In fact, for relatively large $k_l\lambda_{\rm De}$,
kinetic electron trapping plays a key role. In such cases, the
nonlinear evolution in the early-stage is predominately determined
by (i) nonlinear reduction of Landau damping, (ii) nonlinear
frequency downshift of EPW and the accompanied frequency upshift of
the scattered wave, and (iii) pump depletion that becomes important
when the reflectivity rises to a high level; while secondary
instabilities that significantly broaden the $k_l$-spectrum of EPW
and denote the end of the early-stage, mainly consist of trapped
particle instability and generation of beam acoustic modes (cf.
Section~\ref{sec:latestage}). As shown in
Fig.~\ref{Fig:KSpectrumHkLD}(a-f) and
Fig.~\ref{Fig:KSpectrumLkLD}(a-e), broadening of the $k_l$-spectrum
of EPW is sudden and abrupt. Consequently, $t_{\rm early}$ can be
delineated with a great accuracy, yielding a robust $R_{\rm sat,e}$.
Things are different for the three low $k_l\lambda_{\rm De}$ cases
indicated by open squares in Fig.~\ref{Fig:SRSSatLevel_kLD}(b,d),
where the early-stage saturation is predominately caused by pump
depletion, after which broadly featured LDI begins to become important, resulting in broadening of the
$k_l$-spectrum. As shown in Fig.~\ref{Fig:KSpectrumLkLD}(f), in these
cases broadening of the $k_l$ spectrum is much more gradual, making
$t_{\rm early}$ subject to individual choices. However, $R_{\rm
sat,e}$ remains insensitive to individual choices of $t_{\rm early}$
since $R_{\rm sat,e}$ is mainly contributed by the first
reflectivity peak with large width and high amplitude, which is
mainly saturated by the pump depletion and hence always contained
within $t<t_{\rm early}$.

\begin{figure}[!hbtp]
  \centering
  \includegraphics[angle=0,width=0.45\textwidth]{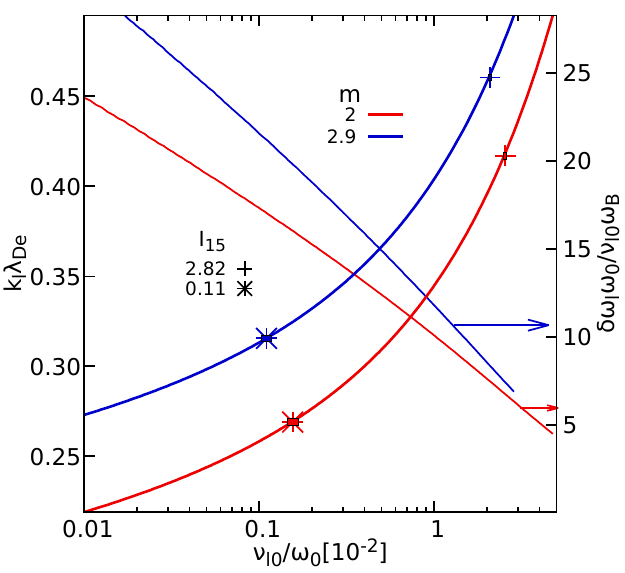}
  \caption{
    Left axis. $k_l\lambda_{\rm De}$ versus $\nu_{l0}$ for $m=2$ (in red) and $m=2.9$ (in blue),
    where $[k_l\lambda_{\rm De}]_{\rm c}$ from the Vlasov-Maxwell simulation is indicated by the plus and asterisk symbols.
    Right axis. $\delta \omega_l\omega_0/\nu_l\omega_{\rm B}$ versus $\nu_l$ for $m=2$ and $m=2.9$, where the nonlinear frequency shift  $\delta \omega_l$ is estimated using Eq.~(\ref{eq:NLOmg_formula}).
  }
  \label{Fig:SRS_Linear}
\end{figure}

In Fig.~\ref{Fig:SRSSatLevel_kLD}(a-b), with decreasing
$k_l\lambda_{\rm De}$, an abrupt rise in $R_{\rm sat,e}$ appears at
one critical $[k_l\lambda_{\rm De}]_{\rm c}$, where the transition
from convective to absolute SRS growth occurs. $[k_l\lambda_{\rm
De}]_{\rm c}$ as obtained from the Vlasov-Maxwell simulations, as
well as the initial Landau damping $\nu_{l0}$ and the convective
gain $G_{\rm R}$ corresponding to $[k_l\lambda_{\rm De}]_{\rm c}$,
is listed in Table~\ref{Tab:absthr} for both $m=2$ and $m=2.9$ at
$I_{15}=2.82$ and $I_{15}=0.11$. It can be seen that due to
nonlinear kinetic effects, $[k_l\lambda_{\rm De}]_{\rm c}$ can far
exceed that from the TWC model, leading to much smaller $[G_{\rm
R}]_{\rm c}$ and hence a sharp rise in the reflectivity at the
transition. Nevertheless, even in the presence of kinetic effects,
$[k_l\lambda_{\rm De}]_{\rm c}$ is larger for a greater $m$. Also
the values of $[\nu_{l0}]_{\rm c}$ corresponding to
$[k_l\lambda_{\rm De}]_{\rm c}$ are quite close for different $m$,
as in the prediction of the TWC model. However, contrary to the TWC
model, in the Vlasov-Maxwell simulation $[\nu_{l0}]_{\rm c}$ is
lower for larger $m$. To comprehend this point, notice that after a
proper account of the nonlinear modification to $\nu_l$ and $\delta
\omega_l$ by kinetic effects,
Eqs.~(\ref{eq:wave3_a0}-\ref{eq:wave3_al}) can still be used to
describe the early-stage evolution of SRS. Various approximations
for the nonlinear evolution of $\nu_l$ and $\delta \omega_l$ have
been proposed in the literature, of which one relatively simple
bounce-averaged model is given in
Ref.~\cite{Yampolsky2009ModelNLLandau,Yampolsky2009NonLinearLandauDamping}
as
\begin{align}
  \nu_l&=\frac{\nu_{l0}}{1+\frac{3\pi^2}{128}\int_0^t\omega_B(t)dt} \label{eq:NLnu_formula}
  \\
  \delta \omega_l&=1.09\omega_B\frac{\omega_{\rm pe}^2}{k_{l}^3(\partial \mathscr{D}_r/\partial\omega_l)}\frac{\partial^2 (f_{e0}/n_{\rm e0})}{\partial v_x^2}|_{v_{\rm phl}}  \label{eq:NLOmg_formula}
\end{align}
where the bouncing frequency $\omega_B=\sqrt{eE_lk_{l}/m_e}$. In
this modified TWC model, when other parameters such as $\gamma_0$,
$v_l$, $v_s$ and $v_0$ are kept nearly the same, the initial Landau
damping $\nu_{l0}$ and $\delta \omega_l/\omega_B$ that depicts the
strength of nonlinear frequency shift, determine the nonlinear
evolution of SRS, and hence whether the absolute SRS growth can
occur or not. Since the increase of either the Landau damping or the
frequency shift would impair the SRS growth, it can be expected that
with increasing $\delta \omega_l/\omega_B$, a lower $\nu_{l0}$ is
required for the onset of absolute SRS growth. This is indeed the
effect of increasing $m$, which leads to greater $\delta
\omega_l/\omega_B$ at the same $\nu_{l0}$, as elucidated in
Fig.~\ref{Fig:SRS_Linear}. Consequently, $[\nu_{l0}]_c$ must be
smaller for increasing $m$ to overcome the effect of stronger
nonlinear frequency shift at greater $m$. It can be further
understood that the different variation of $\delta\omega_l$ and
$\nu_{l0}$ with $m$ ultimately results from $\delta \omega_l \propto
\partial^2f_{\rm e0}/\partial^2 v_x$, in contrast to $\nu_{l0}
\propto \partial f_{\rm e0}/\partial v_x$~\cite{Qiu2021GaussSpec}.
This general understanding should hold even though the simplified
model specified by
Eqs.~(\ref{eq:NLnu_formula}-\ref{eq:NLOmg_formula}) is not precise,
indicating that apart from the Landau damping, the  influence of
super-Gaussian EEDFs on the nonlinear frequency shift is also an
important factor to affect the early-stage development of SRS.

Below $[k_l\lambda_{\rm De}]_{\rm c}$, $R_{\rm sat,e}$ is quite
insensitive to $k_l\lambda_{\rm De}$, except for an intermediate
range of $k_l\lambda_{\rm De}\sim 0.33\text{-}0.4$ for $m=2.9$ and
$k_l\lambda_{\rm De} \sim 0.28\text{-}0.35$ for $m=2$, where $R_{\rm
sat,e}$ increases with decreasing $k_l\lambda_{\rm De}$ as shown in
Fig.~\ref{Fig:SRSSatLevel_kLD}(a). For the high $k_l\lambda_{\rm
De}$ range ($k_l\lambda_{\rm De}>0.4$ for $m=2.9$ and
$k_l\lambda_{\rm De}>0.35$ for $m=2$), the phase velocity of the EPW
is located in the bulk region of the EEDF, leading to strong Landau
damping and also strong electron-trapping induced nonlinearity
(e.g., $\nu_{l0}/\gamma_0\gg 1$ and $\delta \omega_l/\gamma_0\gg
1$). Despite the difference in the initial Landau damping, after the
adjustment period, the absolute SRS growth become similar for
different $k_l\lambda_{\rm De}$ and $m$, until when the reflectivity
reaches the level $\sim 0.02$, where secondary instabilities become
important and the early-stage ends, as shown in
Fig.~\ref{Fig:SRS_early}(a). As a result, the dependence of $R_{\rm
sat,e}$ on both $k_l\lambda_{\rm De}$ and $m$ is much weaken. In the
intermediate range of $k_l\lambda_{\rm De}$ with weaker Landau
damping and kinetic nonlinearity, the Landau damping and the kinetic
nonlinear shift are comparable to $\gamma_0$ (e.g., $\nu_{l0}\sim
0.5\text{-}2.9$ and $\delta \omega_l/\gamma_0\sim 0.3\text{-}1.3$
for $k_l\lambda_{\rm De} \sim 0.28\text{-}0.35$ at $m=2$,
$I_{15}=2.82$ and $\delta n_e/n_{e0}=0.02$).
Thus, the nonlinear adjustment is insufficient to smear the effects of
decreasing initial Landau damping and nonlinear frequency shift when $k_l\lambda_{\rm De}$ decreases or $m$ increases.
Consequently, $R_{\rm sat,e}$ increases with decreasing $k_l\lambda_{\rm De}$ or increasing $m$.
For the low $k_l\lambda_{\rm De}$ range ($k_l\lambda_{\rm De}<0.33$ for $m=2.9$ and $k_l\lambda_{\rm De}<0.28$ for $m=2$) with even smaller initial Landau damping,
after the adjustment period, the Landau damping is nearly negligible.
The absolute growth rate $\gamma_{\rm abs,NL}\lesssim \max[\gamma_{\rm abs}]\equiv 2\gamma_0\sqrt{v_l/v_s}$,
and also the evolution of SRS towards the early-stage saturation,
becomes quite similar for different $k_l\lambda_{\rm De}$ and $m$, as shown in Fig.~\ref{Fig:SRS_early}(b).
As a result,
$R_{\rm sat,e}$ again becomes nearly independent of $k_l\lambda_{\rm De}$ and $m$.

In Fig.~\ref{Fig:SRSSatLevel_kLD}(c-d), it can be seen that the
saturation time $t_{\rm sat}$ is quite large near $[k_l\lambda_{\rm
De}]_{\rm c}$. As shown in Fig.~\ref{Fig:SRS_early}(c), here an
oscillating plateau of the reflectivity is formed before the
absolute growth period, significantly lengthening the adjustment
period.
This is because the countervailing effects of nonlinear Landau damping reduction and the frequency shift induced phase mismatch, nearly balance during the adjustment period.
A slight decrease of $k_l\lambda_{\rm De}$ or increase of $m$,
weakens both the Landau damping and the frequency shift,
thus breaks the balance and significantly reduces the adjustment time,
leading to a sharp drop of $t_{\rm sat}$ with decreasing $k_l\lambda_{\rm De}$ or increasing $m$ near $[k_l\lambda_{\rm De}]_{\rm c}$, as shown in Fig.~\ref{Fig:SRSSatLevel_kLD}(c-d).
This decreasing trend of $t_{\rm sat}$ holds until for $k_l\lambda_{\rm De}$ far below $[k_l\lambda_{\rm De}]_{\rm c}$,
where $t_{\rm sat}$ tends to a value nearly independent of $k_l\lambda_{\rm De}$ and $m$.
Here, the SRS growth is absolute even in the absence of nonlinear kinetic effects, the adjustment period is negligible, and
$t_{\rm sat}$ is primarily contributed by the absolute growth period.
The absolute growth rate $\gamma_{\rm abs}\approx 2\gamma_0\sqrt{v_l/v_s}$
is nearly independent of $k_l\lambda_{\rm De}$ and $m$,
and hence so is $t_{\rm sat}$.

\subsection{The late-stage saturation of SRS}
\label{sec:latestage} Now we examine the impacts of Langdon effect
on the nonlinear saturation of SRS in the late-stage when secondary
instabilities are important. Since the nonlinear behaviour and
dominant saturation mechanism in the late-stage are quite different
between the high $k_l\lambda_{\rm De}$ regime ($0.25\lesssim
k_l\lambda_{\rm De}\lesssim 0.45$) which is investigated under a
high intensity drive ($I_{15}=2.82$), and the low $k_l\lambda_{\rm
De}$ regime ($0.18\lesssim k_l\lambda_{\rm De}\lesssim 0.3$) that is
studied under a low intensity drive ($I_{15}=0.11$), in the
following we discuss them separately.

\begin{figure*}[!hbtp]
  \centering
  \includegraphics[angle=0,width=0.98\textwidth]{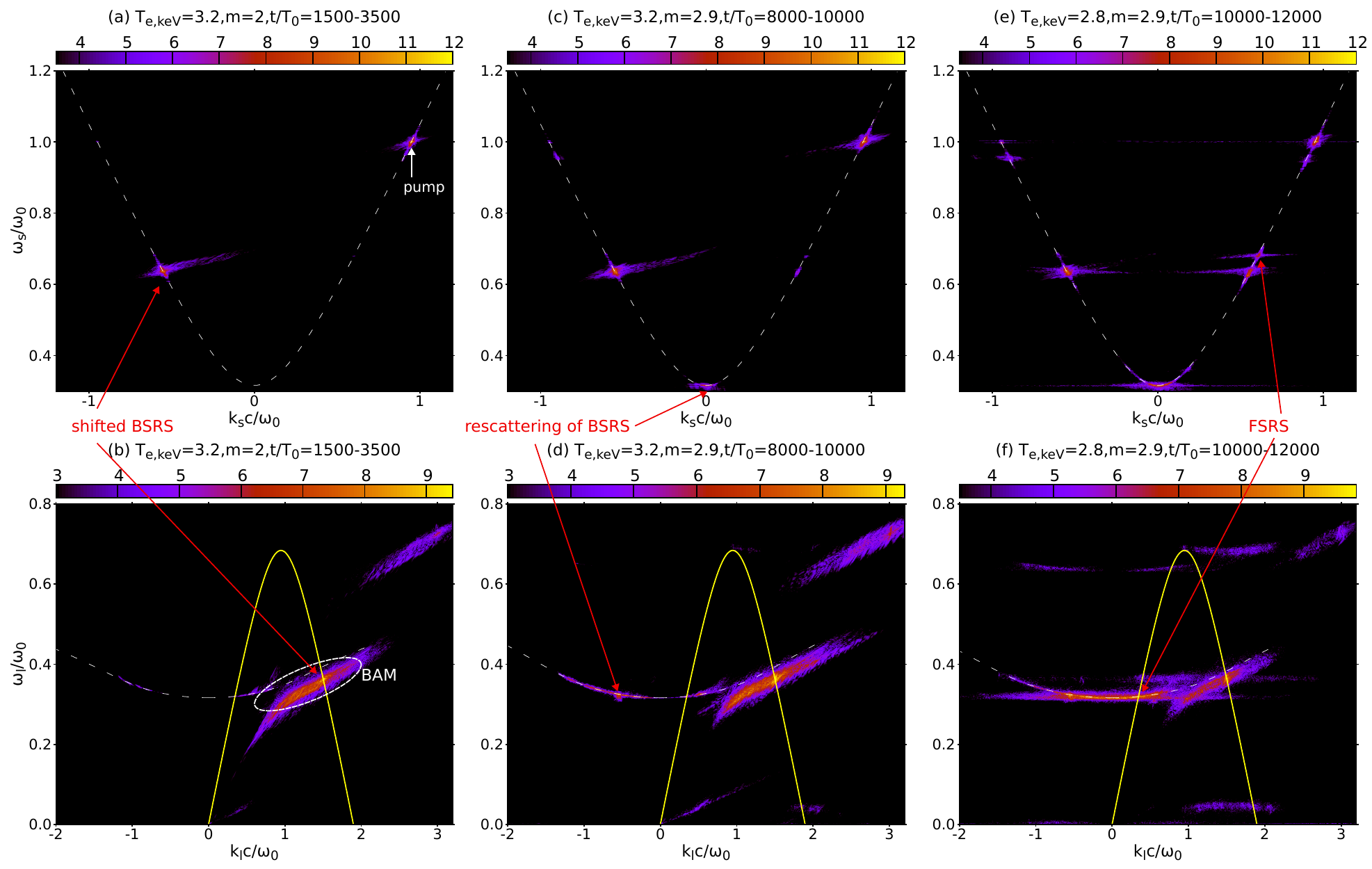}
  \caption{
    Some $\omega$-$k$ spectra of (a,c,e)
    the transverse electric field [$\log_{10}|E_y(\omega,k)|^2$] and (b,d,f) the electrostatic field [$\log_{10}|E_l(\omega_l,k_l)|^2$]
    for cases in Fig.~\ref{Fig:KSpectrumHkLD}.
    The dispersion relations for the EPW and the scattered wave are shown as white dashed lines.
    The Stokes curves, on which $(\omega_0-\omega_l,k_l-k_0)$ satisfies the dispersion relation of the EMW, is plotted as yellow lines.
    The Stokes curves as the locus of the EPW modes that is phase matched for the electromagnetic decay of the pump, i.e., $(\omega_0-\omega_l,k_l-k_0)$ satisfies the dispersion relation of the EMW, is plotted as yellow lines.
The condition $n_e=0.1~n_c$, $\lambda_0=351~\rm nm$ for a homogeneous He plasma with $T_i=T_e/5$ and length of 200$\lambda_0$ is taken.
  }
  \label{Fig:SpeckwHkLD}
\end{figure*}

\begin{figure*}[!hbtp]
  \centering
  \includegraphics[angle=0,width=0.95\textwidth]{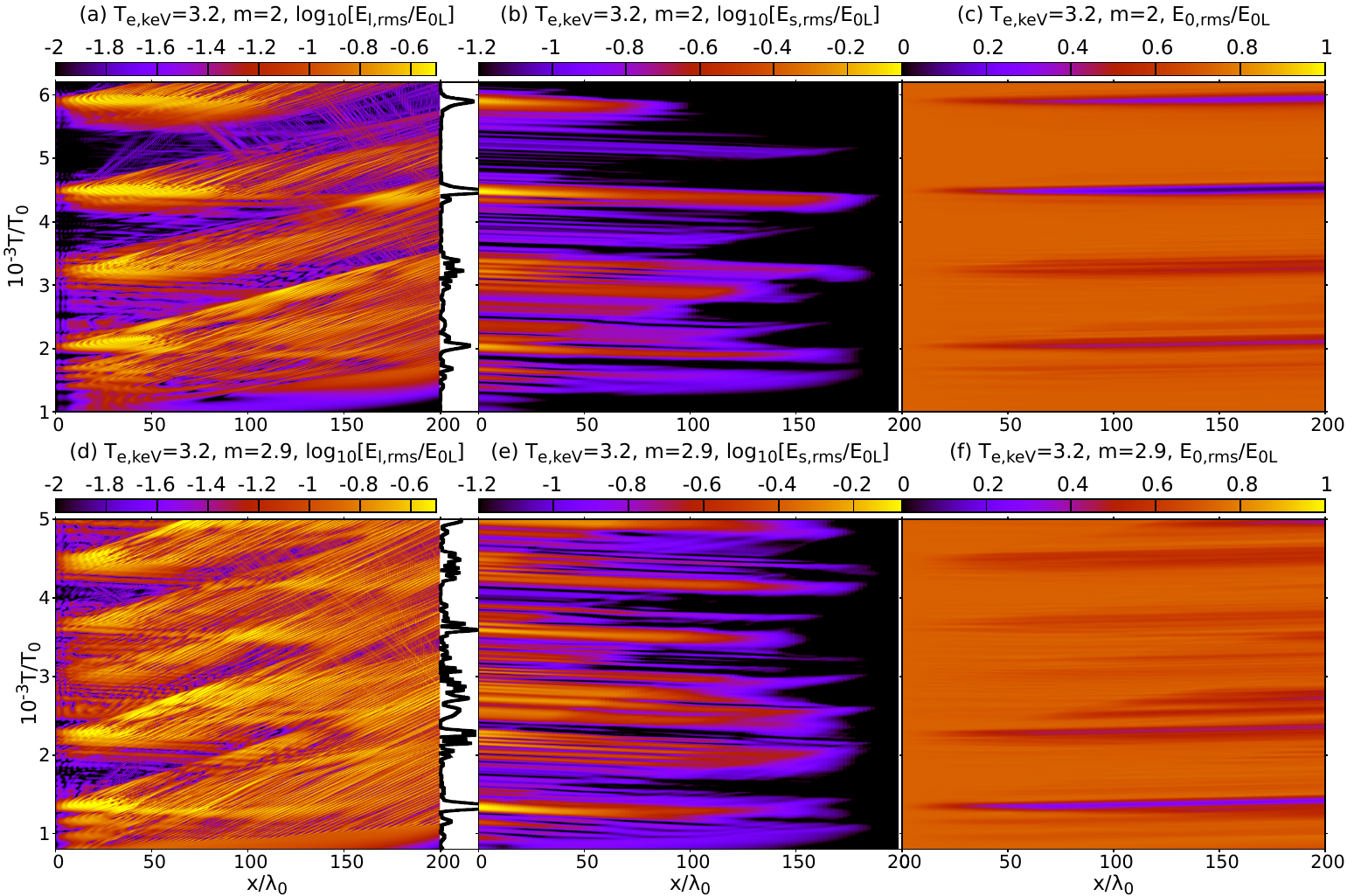}
  \caption{
    The temporal and spatial evolution of (a,d) the electrostatic fields and the transverse electric fields of (b,e) the scattered wave and (c,f) the pump wave for $m=2$ and $m=2.9$.
    The root mean square over one wavelength is taken for each field.
    The reflectivity versus time is shown for comparison as black lines in the right region of panels (a,d).
    The condition $n_e=0.1~n_c$, $\lambda_0=351~\rm nm$ for a homogeneous He plasma with $T_i=T_e/5$ and length of 200$\lambda_0$ is taken.
  }
  \label{Fig:BurstHkLD}
\end{figure*}

For $I_{15}=2.82$ and $T_e=3.2~\rm keV$ as shown in Fig.~\ref{Fig:KSpectrumHkLD}(c-d) for $m=2$ and $m=2.9$ respectively, a burst behaviour of the reflectivity is exhibited.
In the active phase, the trapped particle induced nonlinearity ultimately results in a chaotic state, wherein
a nearly continuous $k_l$-spectrum of the EPW field spreading from $k_{\rm lc}\approx 1.5\omega_0/c$ to about $\omega_0/c$ is generated.
The downward broadening of the wavenumber is caused by vortex-merging processes~\cite{Albrecht-Marc2007VortexMerger,Ghizzo2009HamiltonStochasticSRS},
giving rise to broadband incoherent EPW field consisting of beam acoustic modes~\cite{Yin2006SRSBAM,Yin2006SRSNonMaxBAM,Strozzi2007RamanEAS},
consistent with the nonlinear dispersion relation arising from the EEDF flattening near the phase velocity due to particle trapping.
The decay of the resonant (and usually downshifted) EPW into BAMs breaks the three wave resonance condition,
and can thus suppress the SRS and serve as an efficient saturation mechanism for SRS.
The corresponding typical $k_l\text{-}\omega_l$ spectrum is shown in Fig.~\ref{Fig:SpeckwHkLD}(b), where the BAM feature is obvious.
Note that the adjusted resonant point at $(w_l,k_l)\approx (0.37\omega_0/c,1.50\omega_0/c)$, which corresponds to the intersection of BAMs with the Stokes curve, is downshifted relative the linear resonance mode $(\omega_{lc},k_{lc})=(0.39\omega_0,1.47\omega_0/c)$.
Correspondingly, the backscattered feature in the transverse electric field at $(\omega_s,k_s)\approx (0.63\omega_0,-0.55\omega_0/c)$, as shown in Fig.~\ref{Fig:SpeckwHkLD}(a), is upshifted relative the linear resonance mode $(\omega_{sc},k_{sc})\approx (0.61\omega_0,-0.52\omega_0/c)$.
For $m=2$ where the bursts are well separated,
the temporal-spatial evolution of the EPW, the scattered wave and the pump wave is shown in Fig.~\ref{Fig:BurstHkLD}(a-c).
It can be seen that
the BAMs developed between $1400T_0$ and $1800T_0$ and their convection and damping
lead to oscillation of the reflectivity during this period.
Then at 2000$T_0$, a strong peak occurs.
Accordingly, the laser pump is depleted,
while the decay to BAMs is significantly enhanced by the strong EPW field.
As a result, the reflectivity quickly drops to near zero at $t\sim 2200T_0$.
Then, the pump intensity is restored.
However, the BAM packet needs a much longer time ($\sim L/v_l$) to damp or convect through the plasma~\cite{Fahlen2009WavePacket}.
Inside the packet, the rapid decay to BAMs caused by the large EPW field keeps the (upshifted) backscattered light at a low level.
As this scattered light moves outside the packet into the unperturbed plasma region behind the packet, it is
off-resonance with $\Delta \omega\approx \delta \omega_l$.
This limits its further amplification,
and also
produces a beat pattern separated at $\tau=2\pi/\Delta \omega$~\cite{Winjum2010WavePacketOnSRS};
correspondingly, many high frequency minor bursts appears during the quiescent period between $2500T_0$ and $3000T_0$.
When the packet has almost convected out of the plasma ($t\sim 3200T_0$),
a new major burst coming from new SRS growth in the plasma, now almost clear of the incoherent BAMs, occurs.
Again, the pump depletion that takes effect instantaneously when the reflectivity is large,  and the generation of BAMs whose effects last a long time, begin to suppress the reflectivity.
The continual suppression and recovery of SRS lead to a sequence of major bursts with period about 1300$T_0$,
intervened by many minor bursts.
For $m=2.9$, the decay to BAMs and the pump depletion play similar roles and act as the predominant saturation mechanism for $t_{\rm early}<t<4000T_0$.
Nevertheless,
as shown in Fig.~\ref{Fig:BurstHkLD}(d-f),
due to the smaller Landau damping and hence the greater growth rate and gain of SRS at larger $m$,
significant SRS re-growth can occur behind the packet even when
less than half the plasma is clear of the incoherent BAM field, permitting several packets to coexist and interact.
For example, at $t\sim 2000T_0$, new packet II is formed near the left boundary when the previous packet I has just convected half through the plasma.
The bursts of the reflectivity at $t\sim 2200T_0$ are generated deep inside packet I at $x\approx 150\lambda_0$, and further amplified across packet II on its path to the left boundary, similar to the high gain case in \cite{Winjum2010WavePacketOnSRS}.
So, the bursts become overlapped, 
while the significant interaction between bursts leads to a less regular burst behaviour.
Consequently, compared to $m=2$, the quiescent period is significantly reduced, and the average reflectivity is enhanced.
Besides,
over a long timescale $t>4000T_0$, in addition to decay to BAMs,
rescattering also become important, featured by the remarkable $k_l$-features near $-0.54\omega_0/c$\footnote{The $k$-spectrum itself can not distinguish between the negative and positive $k_l$ components, yet $\omega$-$k$ spectrum can distinguish these two.}.
The typical $\omega_s$-$k_s$ spectrum of the scattered wave,
and the corresponding  $\omega_l$-$k_l$ spectrum of the EPW,
are shown in Fig.~\ref{Fig:SpeckwHkLD}(c) and (d), respectively,
where the feature with $(\omega_l,k_l)\approx (0.32\omega_0,-0.54\omega_0/c)$ and $(\omega_s,k_s)\approx (0.3\omega_0,0)$
is due to rescattering of the primary upshifted backward scattered wave with $(\omega_s,k_s)\approx (0.64\omega_0,-0.54\omega_0/c)$.
Consequently, the SRS becomes more chaotic, leading to more irregular variation of the reflectivity.

For $T_e=2.8~\rm keV$ and $I_{15}=2.82$ as shown in Fig.~\ref{Fig:KSpectrumHkLD}(e-f),
the evolution of SRS after $t>t_{\rm early}$ can be further divided into three periods:
I. In the initial period with $t<4000T_0$, the generation of BAMs plus the pump depletion remains the dominant saturation mechanism.
The SRS reflectivity consists of overlapped bursts for both $m=2$ and $m=2.9$,
and though less apparent, still the quiescent period is shorter for $m=2.9$.
II. In the intermediate period with $4000T_0<t<6000T_0$ for $m=2$ and $4000T_0<t<7000T_0$ for $m=2.9$, rescattering aids in limiting the SRS level.
III. In the final period with $t>6500T_0$ for $m=2$ and $t>8500T_0$ for $m=2.9$,
the decay to low-$k$ Langmuir branch (including modes due to rescattering and forward SRS) is significantly enhanced, leading to the $k_l$-spectrum continuously ranging from $k_{lc}$ down to zero.
As shown in Fig.~\ref{Fig:SpeckwHkLD}(f) for the typical $\omega_l$-$k_l$ spectrum of the EPW,
both BAMs and the low-$k$ Langmuir branch with features due to rescattering and forward SRS contained
are nearly fully occupied.
Correspondingly,
the $\omega_s$-$k_s$ spectrum of the scattered wave shown in Fig.~\ref{Fig:SpeckwHkLD}(e) exhibits features of rescattering with $(\omega_s,k_s)$ around $(0.3\omega_0,0)$
and forward SRS with $(\omega_s,k_s)$ around $(0.68\omega_0/c,0.61\omega_0/c)$.
Such fully developed incoherence results in the drop of SRS reflectivity in the final period,
as demonstrated in Fig.~\ref{Fig:KSpectrumHkLD}(e-f).

\begin{figure*}[!hbtp]
  \centering
  \includegraphics[angle=0,width=0.98\textwidth]{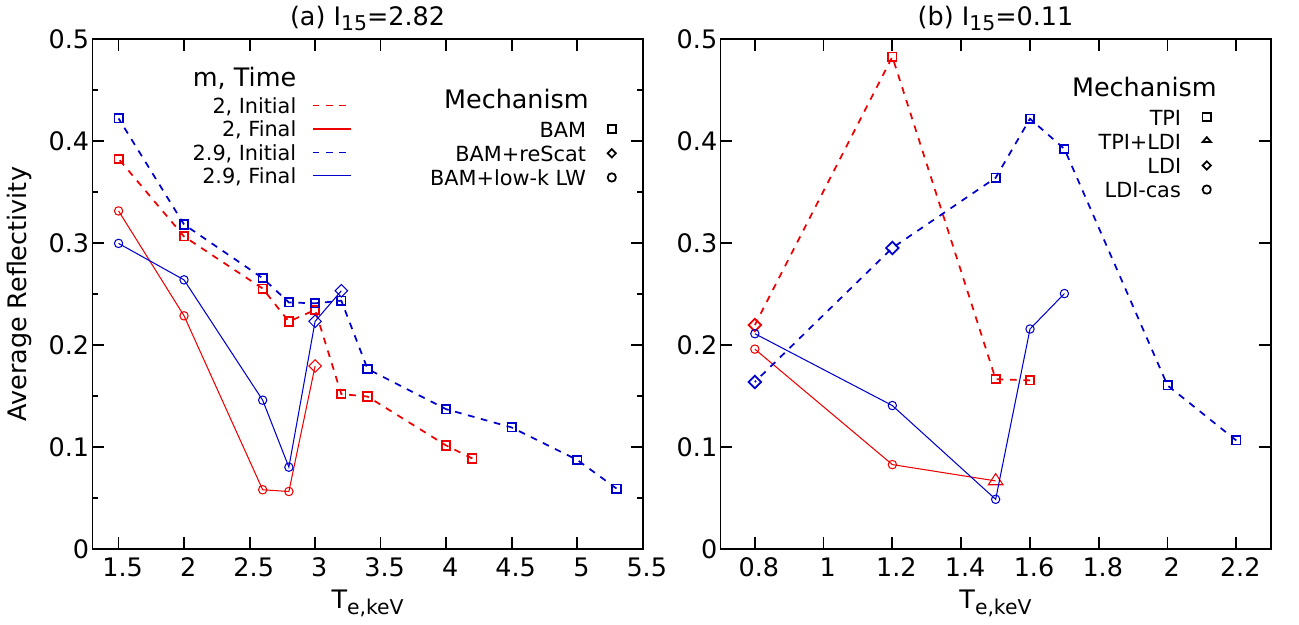}
  \caption{
    The average reflectivity and dominant secondary processes during the late-stage of SRS.
    In (a) with higher $k_l\lambda_{\rm De}$ and $I_{15}=2.82$, three saturation mechanisms are distinguished: the decay to BAMs (open squares), decay to BAMs plus rescattering (open diamonds), and decay to BAMs and low-$k_l$ Langmuir branch (open circles).
    In (b) with lower $k_l\lambda_{\rm De}$ and $I_{15}=0.11$, four saturation mechanisms are distinguished: TPI (open squares), TPI plus LDI (open triangles), LDI (open diamonds) and LDI cascade (open circles).
    When there is a distinct change in the dominant saturation mechanism,
    the average reflectivities are calculated separately
    for the initial period and the final period.
    Notice that in all cases the pump depletion can also play some role in the saturation of SRS, especially at the time with large instantaneous reflectivity.
The condition $n_e=0.1~n_c$, $\lambda_0=351~\rm nm$ for a homogeneous He plasma with $T_i=T_e/5$ and length of 200$\lambda_0$ is taken.
  }
  \label{Fig:SRS_SatMech}
\end{figure*}

The variation of the late-stage saturation mechanism with $T_e$, together with the corresponding average reflectivity, is summarized in Fig.~\ref{Fig:SRS_SatMech}(a) for $m=2$ and $m=2.9$ at $I_{15}=2.82$.
For all cases, initially the dominant saturation mechanism is decay to BAMs plus the pump depletion.
In this period, the lower Landau damping and hence greater gain for increasing $m$ results in shorter quiescent period and hence greater average reflectivity.
With decreasing $T_e$ or increasing $m$,
rescattering and decay to low-$k_l$ Langmuir branch can become important in the later period, usually leading to a more chaotic state with reduced average reflectivity.
Consequently, the dependence of the averaged reflectivity on $m$ becomes more uncertain in the later period,
and the average reflectivity can be smaller for increasing $m$ in some cases.

\begin{figure*}[!hbtp]
  \centering
  \includegraphics[angle=0,width=0.98\textwidth]{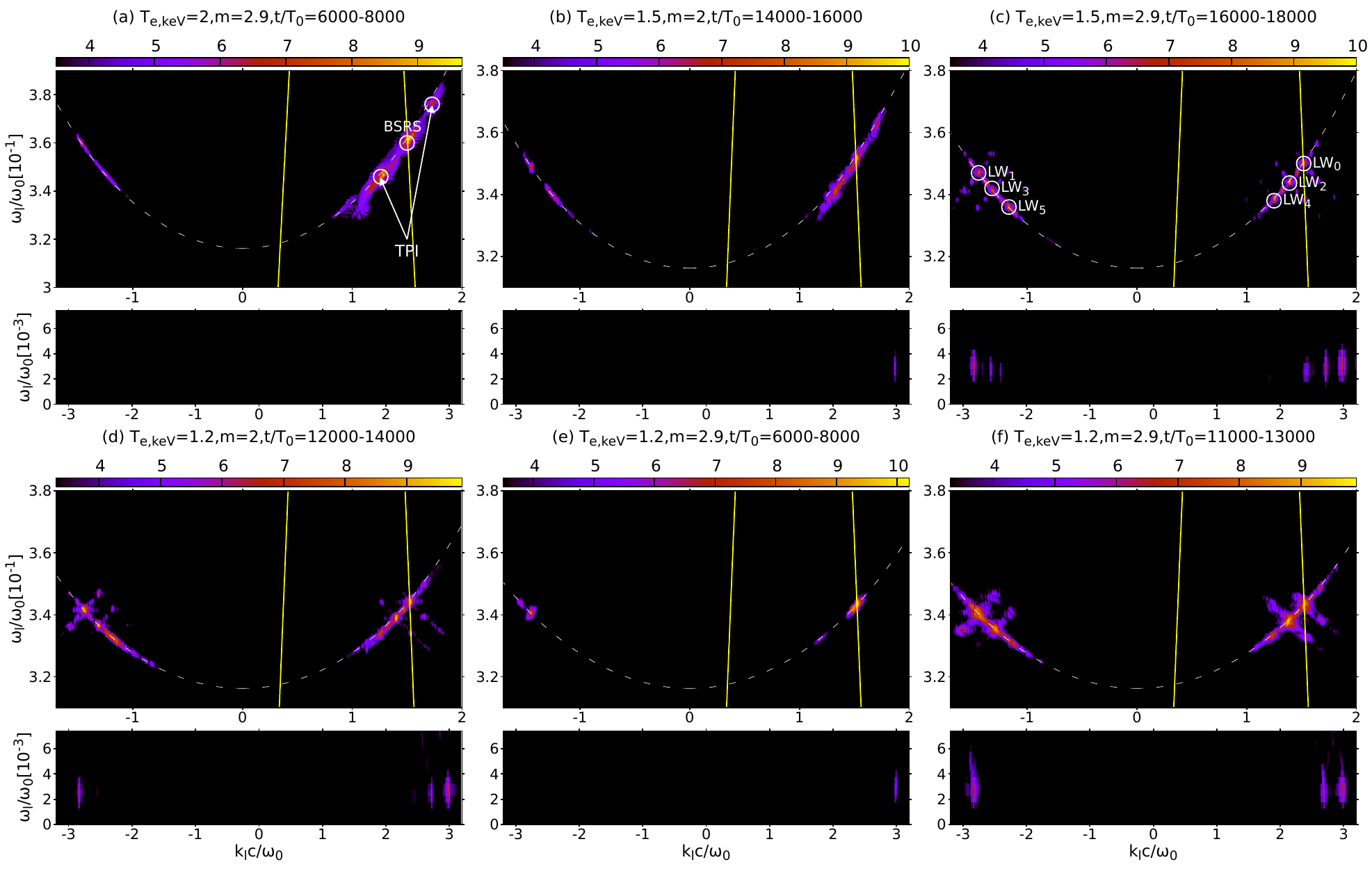}
  \caption{
    Some $\omega_l$-$k_l$ spectra of the electrostatic field [$\rm log_{10}|E_l(\omega_l,k_l)|^2$]
    for cases in Fig.~\ref{Fig:KSpectrumLkLD}.
    In each panel, the upper part shows features due to the EPW,
    while the bottom part demonstrates features due to the LDI generated IAW.
    The dispersion relations for the EPW are shown as white dashed lines.
    The Stokes curves as the locus of the EPW modes that is phase matched for the electromagnetic decay of the pump, i.e., $(\omega_0-\omega_l,k_l-k_0)$ satisfies the dispersion relation of the EMW, is plotted as yellow lines.
The condition $n_e=0.1~n_c$, $\lambda_0=351~\rm nm$ for a
homogeneous He plasma with $T_i=T_e/5$ and length of 200$\lambda_0$
is taken.
  }
  \label{Fig:SpeckwLkLD}
\end{figure*}

For $T_e=1.5~\rm keV$ and $I_{15}=0.11$ as shown in
Fig.~\ref{Fig:KSpectrumLkLD}(c-d), in the initial period after
$t>t_{\rm early}$ ($t<12000T_0$ for $m=2$ and $t<7000T_0$ for
$m=2.9$), the dominant secondary process is trapped particle
instability
(TPI)~\cite{Brunner2004TrapBurstSRS,Brunner2014SideBandInstability,Kruer1969TPI},
featured by the appearance of two sidelobes with $k_l=k_{lc}\pm
\Delta k_{\rm TPI}$ around the primary component $k_{lc}$. The
sidelobe frequency is resonant with the electron bouncing frequency
in the wave frame, thus satisfying $\Delta \omega_{\rm TPI}-\Delta
k_{\rm TPI}v_{\rm phl}=\pm
\omega_B$~\cite{Brunner2014SideBandInstability}, where $\Delta
\omega_{\rm TPI}\equiv \omega_{\rm TPI}-\omega_{lc}$, $\Delta k_{\rm
TPI}=k_{\rm TPI}-k_{lc}$, and $v_{\rm phl}=\omega_{lc}/k_{lc}$ is
the EPW phase velocity. As shown in Fig.~\ref{Fig:SpeckwLkLD}(a) for
the typical $\omega_l$-$k_l$ spectrum of the EPW when TPI dominates,
the most significant mode occurs along the dispersion relation of
EPW, thus $\Delta \omega_{\rm TPI}/\Delta k_{\rm TPI}=v_{l}$, giving
$\Delta k_{\rm TPI}=\pm \omega_B/(v_{\rm phl}-v_{l})$. For
$T_e=1.5~\rm keV$, using $2\pi eE_l/m_e \omega_0 c\approx 0.01$
estimated from the simulation data, it can be obtained $\Delta
k_{\rm TPI}\approx \pm 0.26 \omega_0/c$, consistent with the
sidelobe locations on the $k_l$-spectrum as shown in
Fig.~\ref{Fig:KSpectrumLkLD}(c-d).

As time evolves, Langmuir decay instability~\cite{Russell1999SRSSaturation,Feng2018AntiLangmuirDecayInstability},
where a primary Langmuir wave (LW) decays into a secondary Langmuir wave and an ion acoustic wave (IAW),
begins to become important.
LDI satisfies the matching condition
\begin{equation}
  \begin{aligned}
    \omega_{l,i}&=\omega_{l,i+1}+\omega_a \\
    k_{l,i}&=k_{l,i+1}+k_a
  \end{aligned}
  \label{eq:LDImatch}
\end{equation}
where $i$ denotes the stage number of the Langmuir cascade with $i=0$ corresponding to the primary LW,
and $\omega_a$ and $k_a$ are the frequency and wave number of the IAW, respectively.
The dispersion relation for the LW can be approximated by $\omega_{l,i}^2=\omega_{\rm pe}^2+3k_{l,i}^2v_{\rm the}^2$,
while the dispersion relation for the IAW is approximately $\omega_a=|k_a|c_s$, where $c_s\approx \sqrt{ZT_e/M}$ is the acoustic velocity with $Z$ and $M$ being the charge and mass of the ion species.
Substituting these dispersion relations into Eq.~(\ref{eq:LDImatch}) yields $k_{l,i+1}\approx -k_{l,i}+\Delta k_{\rm LDI}$ and $k_a\approx 2k_{l,i}$,
where the wavenumber difference between two successive cascade step is
\begin{equation}
  \Delta k_{\rm LDI}\approx \frac{2c_s\omega_{l}}{3v_{\rm the}^2}.
  \label{eq:LDIDK}
\end{equation}
The LDI threshold can be estimated by~\cite{Berger1998SRS_SBS}
\begin{equation}
  \frac{\epsilon_0E_{\rm l,LDI}^2}{n_0T_e}=16\frac{\nu_e}{\omega_e}\frac{\nu_a}{\omega_a}
  \label{eq:LDIthreshold}
\end{equation}
where $\nu_a/\omega_a\approx 0.099$ for $T_e/T_i=5$ in He plasma
considered here. For $m=2$ and $T_e=1.5~\rm keV$, when $t\approx
10000T_0$, $E_l/E_{\rm l,LDI}\approx 1.14$, LDI with one cascade
step is excited and results in the reduction of reflectivity in the
later period, as seen from Fig.~\ref{Fig:KSpectrumLkLD}(c). The
corresponding $\omega_l$-$k_l$ spectrum of the electrostatic field
is shown in Fig.~\ref{Fig:SpeckwLkLD}(b), where the features at
$k_l\approx -1.51\omega_0/c+0.07\omega_0/c$ due to the secondary LW
and at $k_l\approx 3\omega_0/c$ due to IAW are obvious. For $m=2.9$,
due to decreasing $\nu_l$ with increasing $m$, $E_{\rm l,LDI}$ is
reduced while $E_l$ is generally greater, leading to $E_l/E_{\rm
l,LDI}\approx 15$ at $t\approx 7000T_0$. Consequently, at least five
LDI cascade steps are apparent from Fig.~\ref{Fig:KSpectrumLkLD}(d)
and Fig.~\ref{Fig:SpeckwLkLD}(c) for $t>10000T_0$, while the
sidelobes of TPI becomes insignificant. This indicates that LDI
cascade becomes the dominant saturation mechanism, limiting the
reflectivity to a very low level about $5\%$ for $t>15000T_0$.

For $T_e=1.2~\rm keV$ and $m=2$ as shown in
Fig.~\ref{Fig:KSpectrumLkLD}(e), in the initial period $t<8000T_0$,
TPI plus the pump depletion is still the dominant saturation
mechanism. However, when $t\sim 8000T_0$, $E_{\rm l}/E_{\rm
l,LDI}\approx 4$, so LDI cascade can be excited and the reflectivity
in the later period is significantly reduced, as shown in
Fig.~\ref{Fig:KSpectrumLkLD}(e) and Fig.~\ref{Fig:SpeckwLkLD}(d).
For $T_e=1.2~\rm keV$ and $m=2.9$ as shown in
Fig.~\ref{Fig:KSpectrumLkLD}(f), $\nu_l/\omega_l\sim 10^{-6}$ is
quite small. As a result, SRS is strongly driven and grows rapidly
at the early time, leading to $E_{\rm l}/E_{\rm l,thr}\approx 100$
and $\delta n_e/n_e\approx 0.06$  at $t=5000T_0$. Broad-featured LDI
is developed, as shown in Fig.~\ref{Fig:SpeckwLkLD}(e) for
$6000T_0<t<8000T_0$. As more cascade steps are excited, the
$k_l$-spectrum is gradually broadened (in contrast to the sudden
broadening of the $k_l$-spectrum in other cases).
When $t\sim 10000T_0$, multiple cascade steps with broad spectral
features have been excited, as shown in
Fig.~\ref{Fig:SpeckwLkLD}(f). In this strongly-driven regime, SRS is
quite turbulent, and the instantaneous reflectivity varies in a wide
range. This leads to a greater average reflectivity in the later
period compared to $m=2$.

The variation of the late-stage saturation mechanism with $T_e$,
together with the corresponding average reflectivity, is summarized
in Fig.~\ref{Fig:SRS_SatMech}(b) for $m=2$ and $m=2.9$ at
$I_{15}=0.11$. Except for the three strongly driven cases with low
$T_e$ ($T_e=0.8~\rm keV$ and $m=2$, $T_e=0.8~\rm keV$ and $m=2.9$,
$T_e=1.2~\rm keV$ and $m=2.9$), initially the dominant mechanism is
TPI plus the pump depletion, while LDI and LDI cascade can develop
over time for some $T_e$ and $m$, significantly reducing the
reflectivity. Increasing $m$ is favorable for excitation of LDI or
LDI cascade since the LDI threshold is reduced while the EPW
amplitude is typically greater before the onset of LDI due to the
lower Landau damping. This can in turn result in a much stronger
drop of the reflectivity in the later period than $m=2$, thus the
reflectivity at $m=2.9$ in the later period can be smaller than
$m=2$.

%
\section{Discussion and summary}
\label{sec:conc} In summary, the influence of Langdon effect on the
nonlinear evolution of SRS over a long timescale is investigated for
a wide range of plasma parameters. 
For the early-stage of SRS, it is found that the Langdon
effect can significantly widen the parameter range for absolute SRS
growth, and the kinetic nonlinear effect can widen this parameter
range further. The time for SRS to reach the early-stage saturation
is significantly reduced by the Langdon effect except when
$k_l\lambda_{\rm De}$ is far below $[k_l\lambda_{\rm De}]_c$. For
the late-stage of SRS, at high $k_l\lambda_{\rm De}$, initially the
dominant saturation mechanism is decay to BAMs plus the pump
depletion, wherein the time-varying reflectivity is composed of a
series of (possibly overlapped and irregular) bursts. The Langdon
effect can shorten the quiescent period and hence increase the
average reflectivity. Additional secondary instabilities such as
rescattering of the primary scattered wave, and the generation of
low-$k$ Langmuir branch can also develop over time, typically
further reducing the reflectivity. The Langdon effect favors the
development of additional secondary processes, though the effect is
generally too weak to make a great difference in the reflectivity.
At low $k_l\lambda_{\rm De}$, the saturation mechanisms include TPI
plus the pump depletion, which dominates in the initial period of
the late-stage except for very low $k_l\lambda_{\rm De}$, and LDI
with single or multiple cascade steps, which typically becomes
important in the later time period. The Langdon effect decreases the
threshold for LDI, thus LDI or LDI cascade with more cascade step is
easier to be excited. This can significantly suppress the
reflectivity at the later time, even leading to smaller reflectivity
than $m=2$. These findings are helpful to comprehend the evolution
behavior of SRS in realistic ICF plasmas, where the prevalent
existence of high intensity speckles and beam overlapping, in
combination with the wide plasma parameter range, can make the
Langdon effect quite important~\cite{Qiu2021GaussSpec}.

\section{acknowledgments}
This work was supported by the National Key R\&D Program of China
(Grant No.~2017YFA0403204), the National Natural Science Foundation
of China (Grant No.~11875093 and~11875091), and the Project
supported by CAEP Foundation (Grant No. CX20210040).
\bibliographystyle{unsrt}
\bibliography{citation}

%
\end{document}